\documentclass[11pt,graphicx,amsmath]{article}
\usepackage{amsmath}
\usepackage{graphicx}
\usepackage{CJK}
\usepackage{bm}
\usepackage{xcolor}
\usepackage[numbers,sort&compress]{natbib}

\def\gsim{\;\rlap{\lower 2.5pt  \hbox{$\sim$}}\raise 1.5pt\hbox{$>$}\;}
\def\lsim{\;\rlap{\lower 2.5pt  \hbox{$\sim$}}\raise 1.5pt\hbox{$<$}\;}
\def\edth{\;\raise1.0pt\hbox{$'$}\hskip-6pt\partial\;}
\def\baredth{\;\overline{\raise1.0pt\hbox{$'$}\hskip-6pt \partial}\;}
\def\be{\begin{equation}}
\def\ee{\end{equation}}
\def\ba{\begin{eqnarray}}
\def\ea{\end{eqnarray}}
\def\nn{\nonumber}

\def\bl#1\el{\begin{align}#1\end{align}}
\def\l{\left}
\def\r{\right}

\title{ Second-order cosmological perturbations. II. Produced by scalar-tensor and tensor-tensor couplings
               }

\author{\small Yang  Zhang  \thanks{yzh@ustc.edu.cn} \  ,
Fei Qin\thanks{chin@mail.ustc.edu.cn}\ ,
Bo Wang  \thanks{ymwangbo@mail.ustc.edu.cn}
         \\
 \small   Department of  Astronomy, Key Laboratory for Researches in Galaxies and Cosmology, \\
 \small    University of Science and Technology of China,   Hefei, Anhui, 230026,  China   }

 \date{}

\begin{document}

\maketitle

\large

\begin{center}
\text{\large\bf Abstract}
\end{center}

We study the  second-order perturbations in the Einstein-de Sitter Universe
in synchronous coordinates.
We solve the second-order  perturbed Einstein equation
with scalar-tensor, and tensor-tensor  couplings between  1st-order perturbations,
and obtain, for each coupling,
the solutions of scalar, vector, and tensor metric perturbations,
  including both the growing and decaying modes
for general initial conditions.
We perform  general synchronous-to-synchronous  gauge transformations up to 2nd order,
which are generated by a 1st-order vector field and a 2nd-order vector field,
and obtain all the residual gauge modes of
the 2nd-order metric perturbations in synchronous coordinates.
We show that only the 2nd-order vector field is effective
for the 2nd-order transformations that we consider
because the 1st-order vector field was already
fixed in obtaining the 1st-order perturbations.
In particular,
the 2nd-order  tensor is invariant
under 2nd-order gauge transformations using  $\xi^{(2)\mu}$ only,
just like the 1st-order tensor is invariant
under 1st-order  transformations.

\


\Large

\section{Introduction}
\

Metric perturbations of Lemaitre-Robertson-Walker spacetimes
within general relativity
are  the theoretical foundation of cosmology.
In the past,  the linear   perturbations of scalar type
\cite{Lifshitz1946,LifshitzKhalatnikov1963,Weinberg1972,Peebles1980,
              PressVishniac1980,Grishchuk1994,MaBertschinger1995,CaiZhang2012}
have been used  in the calculation of cosmic microwave background
(CMB)          and well tested  in  the  measurements of
CMB anisotropies and polarization \cite{WMAP9Bennett2013,Planck2014}.
As predicted by  generic  inflation models,
besides the  scalar  metric perturbation,
 the tensor   perturbation
is also  generated during the inflationary stage
\cite{Grishchuk,FordParker1977GW,Starobinsky,Rubakov,Fabbri,AbbottWise1984,
    Allen1988,Sahni,Giovannini,Tashiro,zhangyang05}.
However, the magnetic polarization $C^{BB}_l$ induced  by the tensor perturbations
\cite{BaskoPolnarev1984, ZaldarriagaHarari1995,Kosowsky1996,
    ZaldarriagaSeljak1997, Kamionkowski1997, ZhaoZhang2006, XiaZhang2008}
has not been detected by the current CMB observations,
and only some  constraint in terms of  the tensor-scalar ratio of metric perturbations
is  given  as  $r < 0.1 $
over very low frequencies $10^{-18} \sim 10^{-16}$Hz
     \cite{WMAP9Bennett2013,Planck2014}.
This  constraint on the ratio
has been inferred from  CMB anisotropies formed at a redshift at $z\sim 1100$ in the matter era,
which is in a rather late stage  of the expanding  Universe.
Furthermore, it has been also based on the formulations of  linear metric perturbations.
On the other hand,
recently LIGO collaboration announced its direct detections
of gravitational waves  emitted  from  binary black holes   \cite{GW150914},
but did not detect  RGW,
only gave  constraints on the spectral energy density of
relic gravitational waves (RGW),
  in a band $10-2000$ Hz \cite{aLIGO O1},
less stringent than that from the CMB measurements.
By   estimations  \cite{TongZhang2009},
it is still possible for  the current LIGO to  detect  RGW
around  frequencies $\sim 10^2 $ Hz
if the running spectral index of the primordial RGW is large.
In regard to these observational constraints
from CMB measurements and LIGO,
one would like to explore other possibilities
that might affect the tensor cosmological perturbation  significantly
 during the course of cosmic expansion.

To the linear level,
the wave equation of RGW depends upon the scale factor $a(\tau)$ only,
and is homogeneous because the anisotropic stress as its source is negligibly small
except for neutrino free-streaming during radiation era \cite{Weinberg2004,MiaoZhang2007}.
Thus,  the other thing that will  affect   RGW
is the   nonlinear couplings of metric perturbations themselves.
To explore their impacts upon RGW,
one  needs to study  the cosmological perturbations  up to 2nd order,
to see how nonlinear gravity changes  the tensor perturbation.
As is known,
in  perturbation formulations,
there are three types of metric perturbations:
scalar, vector, and tensor.
The 2nd-order Einstein equation contains
the couplings of 1st-order metric perturbations
serving as a part of the source for the 2nd-order   perturbations.
For the Einstein-de Sitter mode filled with irrotational dust,
the 1st-order vector metric perturbation can be set to zero
as it is a residual gauge mode.
As a result,   the couplings  of 1st-order metric perturbations
consist of scalar-scalar, scalar-tensor, and tensor-tensor.
So far, the 2nd-order perturbations have found their applications
in detailed calculations of
CMB anisotropies and polarization \cite{PyneCarroll1996, MollerachHarariMatarrese2004},
in the estimation of the non-Gaussianality of primordial perturbation  \cite{Bartolo2010},
and in relic gravitational waves \cite{AnandaClarksonWands2007,Baumann2007}.
In  the literature  \cite{Tomita1967}
 \cite{MatarresePantanoSa'ez1994,Russ1996,Salopek,MalikWands2004}
 \cite{NohHwang2004}
 \cite{Nakamura2003}
  \cite{Bruni97,Matarrese98}
  \cite{Lu2008},
the studies of 2nd-order metric perturbations
have been mostly on the scalar-scalar coupling,
whereas the  couplings involving the 1st-order tensor
have not been sufficiently investigated,
such as the scalar-tensor and the tensor-tensor.
Ref.~\cite{NohHwang2004} derived
the equation of 2nd-order density perturbation with the tensor-tensor coupling.
In our previous work \cite{ScalarScalar},
we have solved  the 2nd-order  perturbed Einstein equation
with the scalar-scalar coupling in the Einstein-de Sitter model,
and obtained all the solutions of
the  2nd-order scalar, vector, and tensor perturbations,
under   general initial conditions.

In this paper we shall extend the study
to the cases  of  scalar-tensor, and tensor-tensor couplings.
We shall derive the corresponding solutions
of 2nd-order scalar, vector, and tensor metric  perturbations
 with general initial conditions.
In addition, we shall perform  2nd-order   gauge transformations,
and identify the residual gauge modes of
the 2nd-order metric perturbations in synchronous coordinates.

In Sec. 2,
we briefly review  the necessary results of 1st-order  perturbations,
which are  used in calculations of the 2nd order  later.

In Sec. 3,  we  split
the 2nd-order perturbed Einstein equations
as the set of  equations of the energy constraint, momentum constraint, and evolution,
each containing the  scalar-tensor, and tensor-tensor couplings, respectively.

In Sec. 4,
     we derive  the solutions of 2nd-order metric perturbations
     with the scalar-tensor  coupling.

In Sec. 5, we obtain the solutions of 2nd-order metric perturbations
  with the  tensor-tensor coupling.

In Sec. 6,     we derive the 2nd-order gauge modes.

We work within the synchronous coordinates,
and, for simple comparisons with  literature,
  use notations mostly as in Refs.~\cite{Matarrese98,ScalarScalar}.
We use a unit in which the speed of light is $c=1$.

\section{   First-Order Perturbations}

In this section,
we  introduce notations and outline the  results of 1st-order  perturbations,
which will be used in later sections.
We consider the   universe filled with the irrotational, pressureless  dust
with the  energy-momentum tensor  $T^{\mu\nu}=\rho U^\mu U^\nu$,
 where  $\rho $ is the mass density,
 $U^\mu=(a^{-1},0,0,0)$ is 4-velocity such that     $U^\mu U_\mu =  -1$.
As in paper I \cite{ScalarScalar},
we take the perturbations of velocity  to be  $U^{(1)\mu} = U^{(2)\mu}= 0 $.
The nonvanishing component is  $T^{00} = a^{-2}\rho$ and $T_{00} = a^{2}\rho$,
where $\rho$ is written as
\be
\rho=\rho^{(0)}\l(1 + \delta^{(1)}+\frac12\delta^{(2)}  \r),
\ee
where $\rho^{(0)}$ is the background density,
$\delta^{(1)}$, $\delta^{(2)}$ are the 1st, 2nd-order density contrasts.
The spatial flat Robertson-Walker (RW) metric in synchronous coordinates
\be \label{18q1}
ds^2=g_{\mu\nu}   dx^{\mu}dx^{\nu}=a^2(\tau)[-d\tau^2
+\gamma_{ij}   dx^idx^j] ,
\ee
where  $\tau$ is conformal time,
 $a(\tau)\propto\tau^2$ for the Einstein-de Sitter model,
$\gamma_{ij}$  is written as
\be\label{eq1}
\gamma_{ij} =\delta_{ij} + \gamma_{ij}^{(1)} + \frac{1}{2} \gamma_{ij}^{(2)}
\ee
where
$\gamma_{ij}^{(1)}$ and  $\gamma_{ij}^{(2)}$
are the 1st- and 2nd-order metric perturbations, respectively.
From (\ref{eq1}), one has $g^{ij}=a^{-2}\gamma^{ij}$ with
$\gamma^{ij}=\delta^{ij} -\gamma^{(1)ij}
-\frac{1}{2}\gamma^{(2)ij}+\gamma^{(1)ik}\gamma^{(1)j}_{k}$,
where   $\delta^{ij}$ is used to  raise the 3-dim spatial indices  of perturbed metric,
such as   $\gamma^{(1)ik}$ and  $\gamma^{(2)ik}$.
We use  the superscripts or subscripts $\mu,\nu$ etc to denote  0, 1, 2,  3,
and   $i,j $ etc to denote  1, 2, or 3.
The perturbed Einstein equation is
\be\label{pertEinstein}
G^{(A)}_{\mu\nu} =8\pi GT^{(A)}_{\mu\nu},
\ee
where $A= 1,2$ denotes the perturbation order,
and we shall study up to 2nd order.
For each  order of  (\ref{pertEinstein}),
the (00) component is the energy constraint,
 $(0i)$ components are the momentum constraints,
and  $(ij)$ components contain the evolution equations.
The set of (\ref{pertEinstein}) are complete to determine
the dynamics of gravitational systems,
and also imply $T^{(A)\mu\nu}\,_{; \, \nu}=0$,
i.e, the conservation of energy and momentum of matter
by the structure of   general relativity.

The first-order metric perturbation $\gamma^{(1)}_{ij}$
can be written as
\be  \label{gqamma1}
\gamma^{(1)}_{ij}=-2\phi^{(1)}\delta_{ij}  +\chi_{ij}^{(1)}\,,
\ee
where  $\phi^{(1)}$ is the trace part of  scalar perturbation,
and $\chi_{ij}^{(1)}$ is  traceless and can be further decomposed into
a scalar and a tensor
\be \label{xqiij1}
\chi_{ij}^{(1)} =D_{ij}\chi^{\parallel(1)}
               +\chi^{\top(1)}_{ij},
\end{equation}
where  $\chi^{\parallel(1)}$ is a scalar function,
$D_{ij} \equiv  \partial_i\partial_j-\frac{1}{3}\delta_{ij}\nabla^2 $,
and  $D_{ij}\chi^{\parallel(1)}$ is the traceless part of the scalar perturbation,
and $\chi^{\top(1)}_{ij}$  is the tensor part,
satisfying the traceless and transverse conditions:
$\chi^{\top(1)i}\, _i=0$, $\partial^i\chi^{\top(1)}_{ij}=0$.
In this paper,
we do not consider the 1st-order vector  perturbation
since  the matter is an irrotational dust.
However, as shall be seen later,
the 2nd-order vector  perturbation
will appear.
Thus,
 the 2nd-order   perturbation    is  written as
\begin{equation} \label{2c1rcerf}
\gamma^{(2)}_{ij}=-2\phi^{(2)}\delta_{ij}+
\chi^{(2)}_{ij}
\end{equation}
with the   traceless part
\be\label{chi2decompose}
\chi^{(2)}_{ij} = D_{ij} \chi^{\parallel(2)}
+\chi^{\perp(2) }_{ij}+\chi^{\top(2)}_{ij}.
\ee
where  the vector mode satisfies a condition
\be\label{chiperpDiver}
\partial^i\partial^j  \chi^{\perp(2) }_{ij}=0,
\ee
which can be written in terms of a curl  vector
\be\label{chiVec1}
\chi^{\perp(2) }_{ij}  = 2 A_{(i, \\j)}  \equiv \partial_i A_j+\partial_j A_i,
\,\,\,\,\,    \partial^i A_i =0.
\ee
 Since the 3-vector $A_i$ is divergenceless and has only two independent components,
  the vector metric perturbation $\chi^{\perp(2) }_{ij}$
has two independent polarization modes, correspondingly.
We remark  that the 2nd-order vector mode $\chi^{\perp(2) }_{ij}$ of
the metric perturbation  is inevitably produced
from the interaction of the 1st-order perturbations
even though  the matter is irrotational dust.

The 1st-order   perturbations  are well known,
and we have calculated the 1st-order perturbations in detail
in our previous work of Ref.~\cite{ScalarScalar}.
In this paper,
we shall list the 1st-order results,
and details can be seen in Ref.~\cite{ScalarScalar}.
The 1st-order density contrast is
\be\label{delta0}
\delta^{(1)} = \frac{\tau^2}{6}   \nabla^2 \varphi
                 +\frac{3 X}{\tau^3} .
~~~~~~
\text{with}
~~~~~~
\nabla^2 \varphi \equiv \frac{6}{ \tau_0^2} \delta_{0g}^{(1)},
\ee
where $ \delta_{0g} ^{(1)}$ is the initial value  of the growing mode
at time $\tau_0$,
$\varphi$ is  the  corresponding gravitational potential.
$\delta_{0}^{(1)}\equiv\frac{\tau_0^2}{6}   \nabla^2 \varphi
                 +\frac{3 X}{\tau_0^3}$
will denote the initial value of $\delta^{(1)}$.
And the solutions of two  scalar perturbations are
\be \label{phi1sol}
\phi^{(1)}({\bf x}, \tau )
=   \frac{5}{3}  \varphi({\bf x})
         + \frac{\tau^2}{18}  \nabla^2 \varphi ({\bf x})
         +\frac{ X({\bf x })}{\tau^3},
\ee
\be\label{Dchi1sol}
D_{ij} \chi^{\parallel(1)}({\bf x}, \tau )
  =  -  \frac{\tau^2}{3} \l(\varphi({\bf x})_{,ij}
  - \frac{1}{3} \delta_{ij} \nabla^2 \varphi ({\bf x})\r)
    -\frac{6\nabla^{-2} D_{ij} X({\bf x })}{\tau^3}   ,
\ee
The 1st-order gravitational wave equation is
\be  \label{eqRGW1}
\chi^{\top(1)''}_{ij}+\frac{4}{\tau}\chi^{\top(1)'}_{ij}
-\nabla^2\chi^{\top(1)}_{ij}=0.
\ee
The solution is
\be  \label{Fourier}
\chi^{\top(1)}_{ij}  ( {\bf x},\tau)= \frac{1}{(2\pi)^{3/2}}
   \int d^3k   e^{i \,\bf{k}\cdot\bf{x}}
    \sum_{s={+,\times}} {\mathop \epsilon
    \limits^s}_{ij}(k) ~ {\mathop h\limits^s}_k(\tau)
       , \,\,\,\, {\bf k}=k\hat{k},
\ee
with two polarization tensors in Eq.(\ref{Fourier})
satisfying
\[
{\mathop \epsilon  \limits^s}_{ij}(k) \delta_{ij}=0,\,\,\,\,
{\mathop \epsilon  \limits^s}_{ij}(k)  k_i=0,\,\,\,
{\mathop \epsilon  \limits^s}_{ij}(k) {\mathop \epsilon  \limits^{s'}}_{ij}(k)
       =2\delta_{ss'}.
\]
During the matter dominant stage the mode is given by
\be\label{GWmode}
{\mathop h\limits^s}_k(\tau ) = \frac{1}{a(\tau)}\sqrt{\frac{\pi}{2}}
   \sqrt{\frac{\tau}{2}}
     \big[\, {\mathop d\limits^s}_1(k)  H^{(1)}_{\frac{3}{2} } (k\tau )
          +{\mathop d\limits^s}_2 (k) H^{(2)}_{\frac{3}{2} } (k\tau ) \big],
\ee
where the  coefficients ${\mathop d\limits^s}_1$, ${\mathop d\limits^s}_2$
are determined by the initial condition during inflation
and by subsequent evolutions
through the reheating, radiation dominant stages
\cite{Grishchuk,zhangyang05}.
Here cosmic processes, such as neutrino  free-streaming
        \cite{Weinberg2004,MiaoZhang2007},
QCD  transition,  and $e^+e^-$ annihilation \cite{WangZhang2008}
   only slightly modify  the amplitude of RGW
and will be  neglected  in  this study.
For RGW generated during inflation  \cite{Grishchuk,Rubakov,Fabbri,AbbottWise1984,
    Allen1988},
the two modes ${\mathop h\limits^s}_k(\tau)$ with  $s= {+,\times}$
are usually assumed to be statistically equivalent,
the superscript $s$ can be dropped.

Thus,   the  1st-order metric perturbation is  given by  \cite{Matarrese98,ScalarScalar}:
\begin{equation}  \label{eq28}
  \gamma^{(1)}_{ij} =-\frac{10}{3}\varphi\delta_{ ij}
  -\frac{\tau^2}{3}\varphi_{,ij}
  -\frac{6 }{\tau^3} \nabla^{-2}X_{,ij}   +\chi^{\top(1)}_{ij},
\end{equation}
which will appear as the coupling terms
in the  equations of the second-order perturbation $\gamma^{(2)}_{ij}$.

\section{ The Second-Order Constraints and Evolution Equations}

According to Ref.~\cite{ScalarScalar},
by using the 2nd-order perturbed Einstein equation
$G^{(2)}_{\mu\nu}=8\pi GT^{(2)}_{\mu\nu}$,
and the 2nd-order   density contrast
\bl\label{density 2order}
\delta^{(2)}=
&
\delta^{(2)}_0
-\frac{1}{2}\gamma^{(2)i}_i
+\frac{1}{2}\gamma^{(2)i}_{0\,i}
+\frac{1}{4}(\gamma^{(1)i}_{i})^2
+\frac{1}{4}(\gamma^{(1)i}_{0i})^2
-\frac{1}{2}\gamma^{(1)i}_i\gamma^{(1)j}_{0j}
+\frac{1}{2}\gamma^{(1)ij}\gamma^{(1)}_{ij}
          \nn\\
& -\frac{1}{2}\gamma^{(1)ij}_0\gamma^{(1)}_{0ij}
-\gamma^{(1)i}_i\delta_0^{(1)}
+\gamma^{(1)i}_{0i}\delta_0^{(1)} ,
\el
following the conservation of energy $T^{0\mu}_{\ \ ;\,\mu}$
with $\delta^{(2)}_0$,
$\gamma^{(1)}_{0\,ij}$,
$\gamma^{(2)}_{0\,ij}$ being the initial values at $\tau_0$,
one has
the 2nd-order energy constraint
involving the couplings $\varphi\chi^{\top(1)}_{ij}$
and $X\chi^{\top(1)}_{ij}$ as:
\be  \label{ieq3asdsd1q}
\frac{2}{\tau}\phi^{(2)'}_{s(t)}
-\frac{1}{3}\nabla^2\phi^{(2)}_{s(t)}
+\frac{6}{\tau^2}\phi^{(2)}_{s(t)}
  -\frac{1}{12} D^{ij}\chi^{\parallel(2) }_{{s(t)},ij}= E_{s(t)} \, ,
\ee
and the momentum constraint:
\be  \label{cfta1q}
2\phi^{(2)'}_{s(t),j}
     +\frac{1}{2} D_{ij}\chi^{\parallel(2)\, ',\,i}_{s(t)}
+\frac12\chi^{\perp (2)',\,i}_{{s(t)}i j}
               =M_{s(t)j} \,,
\ee
and the evolution equation:
\ba\label{afk1q}
&& -(\phi^{(2)''}_{s(t)}
+\frac{4}{\tau}\phi^{(2)'}_{s(t)})\delta_{ij} +\phi^{(2)}_{s(t),i j}
+\frac{1}{2} ( D_{ij}\chi^{||(2)''} _{s(t)}
+\frac{4}{\tau}D_{ij}\chi^{||(2)'}_{s(t)})
 \nn\\
&&
+\frac{1}{2}(\chi^{\perp (2)''}_{s(t)\,ij}
+\frac{4}{\tau}\chi^{\perp (2)'}_{s(t)\,ij})
+\frac{1}{2}(\chi^{\top (2)''}_{s(t)\,ij}
+\frac{4}{\tau}\chi^{\top (2)'}_{s(t)\,ij}
-\nabla^2\chi^{\top(2)}_{s(t)\,i j}
)
\nonumber \\
&&
 -\frac{1}{4}D_{kl}\chi^{||(2),\,kl}_{s(t)}\delta_{ij}
+\frac{2}{3}\nabla^2\chi^{||(2)}_{s(t),\,ij }
-\frac{1}{2}\nabla^2D_{ij}\chi^{||(2)}_{s(t)}
=S_{s(t)\,ij}
    \  .
\ea
where
\bl \label{EST}
E_{s(t)} \equiv  & \frac{5\tau}{18}\chi^{\top(1)' ij}\varphi_{,\,ij}
+\frac{5}{9}\chi^{\top(1) ij}\varphi_{,\,ij}
-\frac{\tau^2}{18}\varphi_{,\,ij}\nabla^2\chi^{\top(1) ij}
-\frac{\tau^2}{36}\chi^{\top(1) ij,\,k}\varphi_{,\,ijk}
 \nn \\
& -\frac{2\tau_0^2}{3\tau^2}\varphi^{,\,ij}\chi^{\top(1)}_{0ij}
 -\frac{2}{\tau^2}\delta_{s(t)\,0}^{(2)}
 +\frac{6}{\tau^2}\phi^{(2)}_{s(t)\,0} \nn \\
&
 +\frac{5}{2\tau^4}\chi^{\top(1)'}_{kl}\nabla^{-2}X^{,kl}
-\frac{1}{\tau^3}\nabla^{-2}X^{,kl}\nabla^2\chi^{\top(1)}_{kl}
\nn\\
&
-\frac{1}{2\tau^3}\chi^{\top(1)}_{km,\,l}\nabla^{-2}X^{,klm}
-\frac{12}{\tau_0^3\tau^2}\chi^{\top(1)}_{0kl}\nabla^{-2}X^{,kl} ,
\el
\bl \label{MsTj}
M_{s(t)j}  \equiv&
\frac{\tau^2}{3}\varphi^{,\,kl}\chi^{\top(1)'}_{kl,\,j}
-\frac{\tau^2}{3}\varphi^{,\,kl}\chi^{\top(1)'}_{jk,\,l}
+\frac{\tau^2}{6}\varphi_{,\,j}^{,kl}\chi^{\top(1)'}_{kl}
- \frac{\tau^2}{6}\chi^{\top(1)'}_{kj}\nabla^2\varphi^{,\,k}
\nn\\
& +\frac{\tau}{3}\varphi^{,\,kl}\chi^{\top(1)}_{kl,\,j}
+\frac{5}{3}\varphi^{,\,k}\chi^{\top(1)'}_{k j}
 \nn \\
&
-\frac{9}{\tau^4}\chi^{\top(1)}_{kl,\,j}\nabla^{-2}X^{,kl}
-\frac{6}{\tau^3}\chi^{\top(1)'}_{jk,\,l}\nabla^{-2}X^{,kl}
+\frac{6}{\tau^3}\chi^{\top(1)' }_{kl,  \, j}\nabla^{-2}X^{,kl}
   \nn\\
&
+\frac{3}{\tau^3}\chi^{\top(1)'}_{kl }\nabla^{-2}X^{,kl}_{,\,j}
 -\frac{3}{\tau^3}\chi^{\top (1)' }_{kj}X^{,k} ,\,
\el
\bl\label{Sstij}
S_{s(t)ij}
\equiv&
-\frac{\tau^2}{6} \varphi^{,kl}\chi^{\top(1)''}_{kl}\delta_{ij}
-\frac{2\tau}{3} \varphi^{,k}_{,i}\chi^{\top(1)'}_{kj}
-\frac{2\tau}{3} \varphi_{,j}^{,k}\chi^{\top(1)'}_{k i}
+\frac{\tau}{3} \chi^{\top(1)'}_{ij} \nabla^2 \varphi
-\frac{\tau}{2} \varphi^{,kl}\chi^{\top(1)'}_{kl}\delta_{ij}
\nn\\
&
+\frac{10}{3}\chi^{\top(1)}_{ij}  \nabla^2 \varphi
+\frac{5}{3}\varphi_{,kl}\chi^{\top(1)kl}\delta_{ij}
-\frac{10}{3}\varphi_{,j}^{,k}\chi^{\top(1)}_{ki}
-\frac{10}{3}\varphi_{,i}^{,k}\chi^{\top(1)}_{kj}
+\frac{10}{3}  \varphi\nabla^2\chi^{\top(1)}_{ij}
\nn\\
&
+\frac{\tau^2}{3} \varphi^{,kl}\chi^{\top(1)}_{ij,\,kl}
+\frac{\tau^2}{3} \varphi^{,kl}\chi^{\top(1)}_{kl,\,ij}
-\frac{\tau^2}{3} \varphi^{,kl}\chi^{\top(1)}_{li,\,jk}
-\frac{\tau^2}{3} \varphi^{,kl}\chi^{\top(1)}_{lj,\,ik}
+5  \varphi^{,\,k}\chi^{\top(1)}_{ij,\,k}
\nn
\\
&
-\frac{5}{3}  \varphi^{,\,k}\chi^{\top(1)}_{ki,j}
-\frac{5}{3}  \varphi^{,\,k}\chi^{\top(1)}_{kj,\,i}
+\frac{\tau^2}{6}  \chi^{\top(1)}_{ij,\,k}\nabla^2 \varphi^{,\,k}
-\frac{\tau^2}{6}\chi^{\top(1)}_{ki,\,j} \nabla^2\varphi^{,k}
- \frac{\tau^2}{6}  \chi^{\top(1)}_{kj,\,i}\nabla^2 \varphi^{,\,k}
\nn\\
&
+\frac{\tau^2}{6} \varphi^{,kl}_{,\,i}\chi^{\top(1)}_{kl,\,j}
+\frac{\tau^2}{6} \varphi^{,kl}_{,j}\chi^{\top(1)}_{kl,\,i}
-\frac{\tau^2}{12} \varphi^{,kml}\chi^{\top(1)}_{km,\,l}\delta_{ij}\nn\\
&-\frac{33}{2\tau^4}\chi^{\top(1)'}_{kl}\nabla^{-2}X^{,kl}\delta_{ij}
-\frac{3}{2\tau^3}\chi^{\top(1)}_{kl,\,m}\nabla^{-2}X^{,klm}\delta_{ij}
-\frac{3}{\tau^3}\chi^{\top(1)''}_{kl}\nabla^{-2}X^{,kl}\delta_{ij}
\nn\\
&
-\frac{9}{\tau^4}X\chi^{\top(1)'}_{ij}
+\frac{18}{\tau^4}\chi^{\top(1)'}_{k i}\nabla^{-2}X^{,k}_{,j}
+\frac{18}{\tau^4}\chi^{\top(1)'}_{kj}\nabla^{-2}X^{,k}_{,i}
\nn\\
&
-\frac{3}{\tau^3}X^{,\,k}\chi^{\top(1)}_{k j,\,i}
-\frac{3}{\tau^3}X^{,\,k}\chi^{\top(1)}_{k i,\,j}
+\frac{3}{\tau^3}X^{,\,k}\chi^{\top(1)}_{ij,\,k}
\nn\\
&
-\frac{6}{\tau^3}\chi^{\top(1)}_{lj,\,ik}\nabla^{-2}X^{,kl}
-\frac{6}{\tau^3}\chi^{\top(1)}_{li,\,jk}\nabla^{-2}X^{,kl}
+\frac{6}{\tau^3}\chi^{\top(1)}_{ij,\,kl}\nabla^{-2}X^{,kl}
\nn\\
&
+\frac{6}{\tau^3}\chi^{\top(1)}_{kl,\,ij}\nabla^{-2}X^{,kl}
+\frac{3}{\tau^3}\chi^{\top(1)}_{kl,\,i}\nabla^{-2}X^{,kl}_{,j}
+\frac{3}{\tau^3}\chi^{\top(1)}_{kl,\,j}\nabla^{-2}X^{,kl}_{,\,i} \ .
\el
The subscript $``s(t)"$ denotes those contributed by
the scalar-tensor coupling.
It is seen that
$E_{{s(t)} } $
contains   the  initial values $\delta^{(2)}_{s(t)0}$, $\phi^{(2)}_{s(t)0}$,
        $ \chi^{\top(1)}_{s(t)0ij}$ etc  at $\tau_0$.
Also we   notice that neither the  tensor    $\chi^{\top (2)}_{s(t)ij}$
nor  the vector    $\chi^{\perp (2)}_{s(t)ij}$
appears in the energy constraint (\ref{ieq3asdsd1q}).
We also observe that $M_{s(t)j}$
on the rhs of  (\ref{cfta1q}) has a  nonvanishing curl,
$\epsilon^{ikj}\partial_k M_{s(t)j} \ne 0$,
and,  to balance that,  a vector perturbation    $ \chi^{\perp(2) }_{s(t)ij} $
must be introduced on the lhs of the equation.
Note that,
$S_{s(t)ij}$ plays a role of source of evolution,
and   the 2nd-order   scalar, vector, and tensor perturbations,
 all  appear in the evolution equation (\ref{afk1q}).

Similarly,
by using the subscript ``T" to denote
the 2nd-order terms contributed by
the tensor-tensor coupling,
one has the energy constraint:
\be  \label{ens26}
\frac{2}{\tau}\phi^{(2)'}_{T}
-\frac{1}{3}\nabla^2\phi^{(2)}_{T}
   +\frac{6}{\tau^2}\phi^{(2)}_{T}
    -\frac{1}{12}  D^{ij} \chi^{\parallel(2) }_{{T},ij}= E_{T},
\ee
the momentum constraint:
\be  \label{mos26}
2\phi^{(2)'}_{T,j}
     +\frac{1}{2} D_{ij}\chi^{\parallel(2)\, ',\,i}_{\,T }
+\frac12\chi^{\perp (2)\,',\,i}_{{T}i j}
               =M_{Tj},
\ee
and the evolution equation:
\ba\label{eq34s26}
&& -(\phi^{(2)''}_{T}
+\frac{4}{\tau}\phi^{(2)'}_{T})\delta_{ij} +\phi^{(2)}_{T,i j}
+\frac{1}{2} ( D_{ij}\chi^{||(2)''} _{T}
+\frac{4}{\tau}D_{ij}\chi^{||(2)'}_{T})
 \nn\\
&&
+\frac{1}{2}(\chi^{\perp (2)''}_{T\,ij}
+\frac{4}{\tau}\chi^{\perp (2)'}_{T\,ij})
+\frac{1}{2}(\chi^{\top (2)''}_{T\,ij}
+\frac{4}{\tau}\chi^{\top (2)'}_{T\,ij}
-\nabla^2\chi^{\top(2)}_{T\,i j}
)
\nonumber \\
&&
 -\frac{1}{4}D_{kl}\chi^{||(2),\,kl}_{T}\delta_{ij}
+\frac{2}{3}\nabla^2\chi^{||(2)}_{T,\,ij }
-\frac{1}{2}\nabla^2D_{ij}\chi^{||(2)}_{T}
=S_{T\,ij}
    \  ,
\ea
where
\bl \label{ET}
E_T \equiv  &-\frac{1}{24}\chi^{\top(1) ' ij}\chi^{\top(1) '}_{ij}
-\frac{2}{3\tau}\chi^{\top(1) ' ij}\chi^{\top(1)}_{ij}
+\frac{1}{6}\chi^{\top(1) ij}\nabla^2\chi^{\top(1)}_{ij}
+\frac{1}{8}\chi^{\top(1) ij,k}\chi^{\top(1)}_{ij,k}
\nonumber  \\
 &-\frac{1}{12}\chi^{\top(1)ij,k}\chi^{\top(1)}_{kj,i}
 -\frac{1}{\tau^2}\chi^{\top(1) ij}\chi^{\top(1)}_{ij}
+\frac{1}{\tau^2}\chi^{\top(1) ij}_0 \chi^{\top(1)}_{0ij}
 -\frac{2}{\tau^2}\delta_{T\,0}^{(2)}
 +\frac{6}{\tau^2}\phi^{(2)}_{T\,0},
\el
\bl\label{MTj}
M_{Tj} \equiv&\chi^{\top(1) ik}(\chi^{\top(1)'}_{kj,\,i}
-\chi^{\top(1)'}_{ki,\,j})-\frac{1}{2}\chi^{\top(1) ik}_{,\,j} \chi^{\top(1)'}_{ik} ,
\el
\bl \label{Sstij1}
S_{Tij} \equiv
&\chi^{\top(1)'k}_{i}\chi^{\top(1)'}_{kj}
-\frac18\chi^{\top(1)'kl}\chi^{\top(1)'}_{kl}\delta_{ij}
+\chi^{\top(1)kl}\chi^{\top(1)}_{li,\,jk}
+\chi^{\top(1)kl}\chi^{\top(1)}_{lj,\,ik}
\nn\\
&
-\chi^{\top(1)kl}\chi^{\top(1)}_{kl,\,ij}
-\chi^{\top(1)kl}\chi^{\top(1)}_{ij,\,kl}
+\chi^{\top(1),\,k}_{l i}\chi^{\top(1),\,l}_{k j}
-\chi^{\top(1)k}_{i,\,l}\chi^{\top(1),\,l}_{j k}
\nn\\
&
-\frac12\chi^{\top(1)kl}_{,\,i}\chi^{\top(1)}_{kl,\,j}
+\frac38\chi^{\top(1)kl,\,m}\chi^{\top(1)}_{kl,\,m}\delta_{ij}
-\frac14\chi^{\top(1)}_{ml,\,k}\chi^{\top(1)m k,\,l}\delta_{ij}
\nn
\\
&
+\frac12\chi^{\top(1)kl}\chi^{\top(1)''}_{kl}\delta_{ij}
+\frac{2}{\tau}\chi^{\top(1)kl}\chi^{\top(1)'}_{kl}\delta_{ij}
,
\el
where $E_{T } $
contains   the  initial values $\delta^{(2)}_{T0}$, $\phi^{(2)}_{T0}$,
        $ \chi^{\top(1)}_{T0ij}$ etc  at $\tau_0$.

In the following,
we shall solve  the   set  of equations with scalar-tensor  couplings,
and tensor-tensor couplings respectively.

\section{ 2nd Order   Perturbations
             with the Source  $\varphi\chi^{\top(1)}_{ij}$}

\subsection{Scalar Perturbation  $\phi^{(2)}_{s(t)}$   }

Combining  the constraint equations
[Eq.(\ref{ieq3asdsd1q}) + $\frac{1}{6}\partial^j \int_{\tau_0} ^\tau d\tau' $ Eq.(\ref{cfta1q})]
 gives
\[
\frac{2}{\tau}\phi^{(2)'}_{s(t)}
+\frac{6}{\tau^2}\phi^{(2)}_{s(t)}
          =E_{s(t)}
+\frac16\int_{\tau_0}^\tau d\tau' M_{{s(t)}j}^{,j}
+\frac13\nabla^2\phi^{(2)}_{{s(t)}0}
+\frac{1}{18}\nabla^2\nabla^2\chi^{\parallel(2)\, }_{{s(t)}0} \,.
\]
Substituting the known $E_{s(t)}$ and $M_{s(t)j}^{, j }$ into the above,
using the 1st-order GW equation (\ref{eqRGW1})
to replace $\nabla^2\chi^{\top(1)}_{ij}$ contained in $M_{s(t)j}^{, j }\,$,
one has the first-order differential equation of
$ \phi^{(2)}_{s(t)}$ as the following:
\bl\label{phi2stEq}
&
\phi^{(2)'}_{s(t)}
+\frac{3}{\tau}\phi^{(2)}_{s(t)}
\nn\\
          =&
\frac{\tau^2}{9}\varphi^{,\,kl}\chi^{\top(1)'}_{ kl}
+\frac{\tau}{3}\chi^{\top(1)}_{kl}\varphi^{,\,kl}
+\frac{1}{\tau}\bigg( 3\phi^{(2)}_{s(t)\,0}
 -\delta_{s(t)\,0}^{(2)}
-\frac{\tau_0^2}{3}\varphi^{,\,ij}\chi^{\top(1)}_{0ij}
\bigg)
-\frac{\tau}{12}C
\nn\\
&
+\frac{5}{4\tau^3}\chi^{\top(1)'}_{kl}\nabla^{-2}X^{,kl}
-\frac{6}{\tau_0^3\tau}\chi^{\top(1)}_{0kl}\nabla^{-2}X^{,kl}
+\frac{3\tau}{4}\Big[\nabla^{-2}X^{,\,kl}\Big]\int_{\tau_0}^\tau
\frac{1}{\tau^{'4}}\nabla^2\chi^{\top(1)}_{kl}d\tau' \, ,
\el
where the constant
\bl\label{C}
C\equiv&
\frac{\tau_0^2}{3}\varphi^{,\,kl}\nabla^2\chi^{\top(1)}_{0kl}
+\frac{\tau_0^2}{6}\chi^{\top(1)}_{0kl,m}\varphi^{,\,klm}
-\frac{\tau_0}{3}\varphi^{,\,kl}\chi^{\top(1)' }_{0kl}
+\frac{2}{3}\varphi^{,\,kl}\chi^{\top(1)}_{0kl}
-2\nabla^2\phi^{(2)}_{{s(t)}0}
\nn\\
&
-\frac{1}{3}\nabla^2\nabla^2\chi^{\parallel(2)\, }_{{s(t)}0}
+ \frac{6}{\tau_0^3}\nabla^2\chi^{\top(1) }_{0kl}\nabla^{-2}X^{,kl}
+\frac{3}{\tau_0^3}\chi^{\top(1)}_{0kl ,m}\nabla^{-2}X^{,klm}
      \  ,
\el
depending on the initial values of metric perturbations at $\tau_0$.
The solution of Eq.(\ref{phi2stEq}) is
\bl\label{phi2st}
\phi^{(2)}_{s(t)}
=&
 ( \phi^{(2)}_{s(t)\,0}
 -\frac{1}{3}\delta_{s(t)\,0}^{(2)}
-\frac{\tau_0^2}{9}\varphi^{,\,ij}\chi^{\top(1)}_{0ij}
)
-\frac{\tau^2}{60}C
+\frac{\tau^2}{9}\varphi^{,kl}\chi^{\top(1) }_{kl}
-\frac{2}{9\tau^3}\varphi^{,kl}\int^{\tau}_{\tau_0}
    \tau^{'4}\,\chi^{\top(1) }_{kl} d\tau'
\nn\\
&
+\frac{3\tau^2}{20}(\nabla^{-2}X^{,kl})
    \int_{\tau_0}^{\tau}
    \frac{1}{\tau^{'4}}\nabla^2\chi^{\top(1)}_{kl}d\tau'
-\frac{3}{20\tau^3}(\nabla^{-2}X^{,kl})
    \int^\tau_{\tau_0}\tau'\nabla^2\chi^{\top(1)}_{kl}d\tau'
\nn\\
&
+\frac{5}{4\tau^3}\chi^{\top(1)}_{kl}\nabla^{-2}X^{,kl}
-\frac{2}{\tau_0^3}\chi^{\top(1)}_{0kl}\nabla^{-2}X^{,kl}
 +\frac{W({\bf x})}{\tau^3} \, ,
\el
where   integration by parts has been  used,
and $W({\bf x})$ is a time-independent function.
By letting  $\phi^{(2)}_{s(t)}(\tau_0)=\phi^{(2)}_{s(t)\,0}$ at $\tau=\tau_0$
in  (\ref {phi2st}),
$W({\bf x})$ is fixed as following
\bl\label{W2STdecayMode}
W({\bf x})
 =&
\frac{3}{4}\chi^{\top(1)}_{0kl}\nabla^{-2}X^{,kl}
+\frac{\tau_0^3}{3}\delta_{s(t)\,0}^{(2)}
+\frac{\tau_0^5}{60}C      .
\el
As we have checked,
the solution (\ref{phi2st}) can be  also derived by
 the trace part of the evolution equation (\ref{afk1q})
  together with
the energy constraint (\ref{ieq3asdsd1q}).

\subsection{  Scalar Perturbation $\chi^{||(2)}_{s(t)}$}

   The expression
$\partial^j \int_{\tau_0} ^\tau d\tau' $ Eq.(\ref{cfta1q}) gives
\bl\label{mosInt}
2\nabla^2\phi^{(2)}_{s(t)}
 +\frac{1}{2}   D_{ij}\chi^{\parallel(2),\, ij }_{s(t)}
 =\int_{\tau_0}^\tau d\tau' M_{s(t)j}^{,j}
+2\nabla^2\phi^{(2)}_{s(t)0}
+ \frac{1}{2}   D_{ij}\chi^{\parallel(2),\, ij }_{s(t)0} \, .
\el
Substituting $M_{s(t)j}$ of (\ref{MsTj}) and
$\phi^{(2)}_{s(t)}$ of Eq.(\ref{phi2st})
into the above  yields
\bl\label{chi2st||1}
\chi^{\parallel(2)\, }_{s(t)}
=&
Z+\frac{\tau^2}{10}\nabla^{-2}C
+\nabla^{-2}\bigg[-\frac{2\tau^2}{3}\varphi^{,\,kl}\chi^{\top(1) }_{kl}
+\frac{4}{3\tau^3}\varphi^{,\,kl}\int^{\tau}_{\tau_0}
    \tau^{'4}\,\chi^{\top(1) }_{kl} d\tau'
\bigg]
\nn\\
&
+\nabla^{-2}\nabla^{-2}\bigg[
\tau^2\varphi^{,\,kl}\nabla^2\chi^{\top(1)}_{kl}
+\frac{\tau^2}{2}\chi^{\top(1)}_{kl,m}\varphi^{,\,klm}
-\tau\varphi^{,\,kl}\chi^{\top(1)' }_{kl}
+2\varphi^{,\,kl}\chi^{\top(1)}_{kl}
\bigg]
\nn\\
&
+\nabla^{-2}
\bigg[
-\frac{15}{2\tau^3}\chi^{\top(1)}_{kl}\nabla^{-2}X^{,kl}
-\frac{9\tau^2}{10}(\nabla^{-2}X^{,kl})
    \int_{\tau_0}^{\tau}
    \frac{1}{\tau^{'4}}\nabla^2\chi^{\top(1)}_{kl}d\tau'
\nn\\
&
+\frac{9}{10\tau^3}(\nabla^{-2}X^{,kl})
    \int^\tau_{\tau_0}\tau'\nabla^2\chi^{\top(1)}_{kl}d\tau'
\bigg]
+\nabla^{-2}\nabla^{-2}
\bigg[
\frac{36}{\tau^3}\nabla^2\chi^{\top(1) }_{kl}\nabla^{-2}X^{,kl}
\nn\\
&
+\frac{18}{\tau^3}\chi^{\top(1)}_{kl ,m}\nabla^{-2}X^{,klm}
+(\nabla^{-2}X^{,kl})\int_{\tau_0}^\tau \frac{54}{\tau^{'4}}\nabla^2\chi^{\top(1)}_{kl}d\tau'
\bigg]
-\frac{6}{\tau^3}\nabla^{-2}W \, .
\el
where the constant
\bl
Z\equiv&
\chi^{\parallel(2)\, }_{{s(t)}0}
+\nabla^{-2}
\bigg(
2\delta_{s(t)\,0}^{(2)}
+\frac{2\tau_0^2}{3}\varphi^{,\,ij}\chi^{\top(1)}_{0ij}
\bigg)
+\nabla^{-2}\nabla^{-2}\bigg(
-\tau_0^2\varphi^{,\,kl}\nabla^2\chi^{\top(1)}_{0kl}
\nn\\
&
-\frac{\tau_0^2}{2}\chi^{\top(1)}_{0kl,m}\varphi^{,\,klm}
+\tau_0\varphi^{,\,kl}\chi^{\top(1)' }_{0kl}
-2\varphi^{,\,kl}\chi^{\top(1)}_{0kl}
\bigg)
+\nabla^{-2}\bigg[
\frac{12}{\tau_0^3}\chi^{\top(1)}_{0kl}\nabla^{-2}X^{,kl}
\bigg]
\nn\\
&
+\nabla^{-2}\nabla^{-2}
\bigg[
-\frac{36}{\tau_0^3}\nabla^2\chi^{\top(1) }_{0kl}\nabla^{-2}X^{,kl}
-\frac{18}{\tau_0^3}\chi^{\top(1)}_{0kl ,m}\nabla^{-2}X^{,klm}
\bigg] \, ,
\el
depending on the initial values of metric perturbations at $\tau_0$.
Thus,  the scalar perturbation   $D_{ij}\chi^{||(2) }_{s(t)} $
is obtained
\bl\label{chi112qqq}
D_{ij}\chi^{\parallel(2)\, }_{s(t)}
=&
D_{ij}Z+\frac{\tau^2}{10}D_{ij}\nabla^{-2}C
+D_{ij}\nabla^{-2}\bigg[-\frac{2\tau^2}{3}\varphi^{,\,kl}\chi^{\top(1) }_{kl}
+\frac{4}{3\tau^3}\varphi^{,\,kl}\int^{\tau}_{\tau_0}
    \tau^{'4}\,\chi^{\top(1) }_{kl} d\tau'
\bigg]
\nn\\
&
+D_{ij}\nabla^{-2}\nabla^{-2}\bigg[
\tau^2\varphi^{,\,kl}\nabla^2\chi^{\top(1)}_{kl}
+\frac{\tau^2}{2}\chi^{\top(1)}_{kl,m}\varphi^{,\,klm}
-\tau\varphi^{,\,kl}\chi^{\top(1)' }_{kl}
+2\varphi^{,\,kl}\chi^{\top(1)}_{kl}
\bigg]
\nn\\
&
+D_{ij}\nabla^{-2}
\bigg[
-\frac{15}{2\tau^3}\chi^{\top(1)}_{kl}\nabla^{-2}X^{,kl}
-\frac{9\tau^2}{10}(\nabla^{-2}X^{,kl})
    \int_{\tau_0}^{\tau}
    \frac{1}{\tau^{'4}}\nabla^2\chi^{\top(1)}_{kl}d\tau'
\nn\\
&
+\frac{9}{10\tau^3}(\nabla^{-2}X^{,kl})
    \int^\tau_{\tau_0}\tau'\nabla^2\chi^{\top(1)}_{kl}d\tau'
\bigg]
+D_{ij}\nabla^{-2}\nabla^{-2}
\bigg[
\frac{36}{\tau^3}\nabla^2\chi^{\top(1) }_{kl}\nabla^{-2}X^{,kl}
 \nn\\
&
+\frac{18}{\tau^3}\chi^{\top(1)}_{kl ,m}\nabla^{-2}X^{,klm}
+(\nabla^{-2}X^{,kl})\int_{\tau_0}^\tau \frac{54}{\tau^{'4}}\nabla^2\chi^{\top(1)}_{kl}d\tau'
\bigg]
-\frac{6}{\tau^3}D_{ij}\nabla^{-2}W \,   .
\el
We remark
 that the solution (\ref{chi112qqq}) can be also obtained  by
the traceless part of the evolution equation (\ref{afk1q})  together with
the momentum constraint (\ref{cfta1q}).
Our result (\ref{chi112qqq})  contains
the nonzero initial values (through $\tau_0$) and decaying modes,
and applies to general  situations.

\subsection{Vector Perturbation  $\chi^{\perp(2)}_{s(t)i j}$  }

The time integral of the momentum constraint (\ref{cfta1q})
from $\tau_0$ to $\tau$ is
\be\label{mosInt2}
2\phi^{(2)}_{s(t),j}
     +\frac{1}{3}\nabla^2\chi^{\parallel(2)}_{s(t),j}
+\frac12\chi^{\perp (2),\,i}_{{s(t)}i j}
               =\int_{\tau_0}^{\tau}d\tau' M_{s(t)j}
               +2\phi^{(2)}_{s(t)0,j}
     +\frac{1}{3}\nabla^2\chi^{\parallel(2)}_{s(t)0,j}
+\frac12\chi^{\perp (2),\,i}_{{s(t)}0i j}
\ ,
\ee
Using $M_{s(t)j}$ of (\ref{MsTj})   and  Eq.(\ref{eqRGW1}),
one has
\bl\label{MsTjIntTau0}
\int_{\tau_0}^\tau d\tau' M_{s(t)j}
=&
\int_{\tau_0}^{\tau}\bigg[
-\frac{\tau'}{3}\partial_j(\varphi^{,\,kl}\chi^{\top(1)}_{kl})
+\frac{\tau'}{3}\nabla^2(\varphi^{,\,k}\chi^{\top(1)}_{kj})
\bigg]d\tau'
+\frac{\tau^2}{6}\partial_j(\varphi^{,\,kl}\chi^{\top(1)}_{kl})
\nn\\
&
-\frac{\tau^2}{6}\chi^{\top(1)}_{kj}\nabla^2\varphi^{,\,k}
-\frac{\tau^2}{3}\varphi^{,\,kl}\chi^{\top(1)}_{kj,\,l}
+\frac{\tau^2}{6}\varphi^{,\,kl}\chi^{\top(1)}_{kl,j}
-\frac{\tau}{3}\varphi^{,k}\chi^{\top(1)' }_{kj}
\nn\\
&
+\frac{2}{3}\varphi^{,\,k}\chi^{\top(1)}_{k j}
+\bigg[
-\frac{\tau_0^2}{6}\partial_j(\varphi^{,\,kl}\chi^{\top(1)}_{0kl})
+\frac{\tau_0^2}{6}\chi^{\top(1)}_{0kj}\nabla^2\varphi^{,\,k}
\nn\\
&
+\frac{\tau_0^2}{3}\varphi^{,\,kl}\chi^{\top(1)}_{0kj,\,l}
-\frac{\tau_0^2}{6}\varphi^{,\,kl}\chi^{\top(1)}_{0kl,j}
+\frac{\tau_0}{3}\varphi^{,k}\chi^{\top(1)' }_{0kj}
-\frac{2}{3}\varphi^{,k}\chi^{\top(1)}_{0k j}
\bigg]
\nn\\
&
+\int_{\tau_0}^\tau
\bigg[
\partial_j
\Big(
\frac{3}{\tau^{'3}}\chi^{\top(1)' }_{kl}\nabla^{-2}X^{,kl}
\Big)
-\frac{6}{\tau^{'3}}\chi^{\top(1)'}_{jk,\,l}\nabla^{-2}X^{,kl}
-\frac{3}{\tau^{'3}}\chi^{\top (1)' }_{kj}X^{,k}
\bigg]d\tau'
\nn\\
&
+\frac{3}{\tau^{3}}\chi^{\top(1) }_{kl,  \, j}\nabla^{-2}X^{,kl}
-\frac{3}{\tau_0^{3}}\chi^{\top(1) }_{0kl,  \, j}\nabla^{-2}X^{,kl}
           \ .
\el
Plugging the solutions of $\phi^{(2)}_{s(t)}$ of (\ref{phi2st})
and $\chi^{\parallel(2)}_{s(t)}$ of (\ref{chi112qqq})
into (\ref{mosInt2}),
we directly read $\chi^{\perp (2),\,i}_{{s(t)}i j}$ as
\bl\label{chiSTperp}
\chi^{\perp (2),\,i}_{{s(t)}i j}
=&
Q_j
+\int_{\tau_0}^{\tau}\bigg[
-\frac{2\tau'}{3}\partial_j(\varphi^{,\,kl}\chi^{\top(1)}_{kl})
+\frac{2\tau'}{3}\nabla^2(\varphi^{,\,k}\chi^{\top(1)}_{kj})
\bigg]d\tau'
+\frac{\tau^2}{3}\partial_j(\varphi^{,\,kl}\chi^{\top(1)}_{kl})
\nn\\
&
-\frac{\tau^2}{3}\chi^{\top(1)}_{kj}\nabla^2\varphi^{,\,k}
-\frac{2\tau^2}{3}\varphi^{,\,kl}\chi^{\top(1)}_{kj,\,l}
+\frac{\tau^2}{3}\varphi^{,\,kl}\chi^{\top(1)}_{kl,j}
-\frac{2\tau}{3}\varphi^{,\,k}\chi^{\top(1)' }_{kj}
+\frac{4}{3}\varphi^{,\,k}\chi^{\top(1)}_{k j}
\nn\\
&
+\nabla^{-2}\partial_j\bigg[
-\frac{2\tau^2}{3}\varphi^{,\,kl}\nabla^2\chi^{\top(1)}_{kl}
-\frac{\tau^2}{3}\chi^{\top(1)}_{kl,m}\varphi^{,\,klm}
+\frac{2\tau}{3}\varphi^{,\,kl}\chi^{\top(1)' }_{kl}
-\frac{4}{3}\varphi^{,\,kl}\chi^{\top(1)}_{kl}
\bigg]
\nn\\
&
+ \int_{\tau_0}^\tau
\bigg[ -\frac{12}{\tau^{'3}}\chi^{\top(1)'}_{jk,\,l}\nabla^{-2}X^{,kl}
-\frac{6}{\tau^{'3}}\chi^{\top (1)' }_{kj}X^{,k}
\nn\\
&
+\partial_j\nabla^{-2}
\Big(
\frac{12}{\tau^{'3}}\chi^{\top(1)'}_{kl,m}\nabla^{-2}X^{,klm}
+\frac{6}{\tau^{'3}}\chi^{\top (1)' }_{kl}X^{,kl}
\Big)
\bigg]d\tau'
+\frac{6}{\tau^{3}}\chi^{\top(1) }_{kl,  \, j}\nabla^{-2}X^{,kl}
\nn\\
&
-\partial_j\nabla^{-2}
\bigg(
\frac{6}{\tau^{3}}\nabla^2\chi^{\top(1) }_{kl}\nabla^{-2}X^{,kl}
+\frac{6}{\tau^{3}}\chi^{\top(1) }_{kl, m}\nabla^{-2}X^{,klm}
\bigg) \ ,
\el
where the constant vector
\bl
Q_j\equiv&
\chi^{\perp (2),\,i}_{{s(t)}0i j}
-\frac{\tau_0^2}{3}\partial_j(\varphi^{,\,kl}\chi^{\top(1)}_{0kl})
+\frac{\tau_0^2}{3}\chi^{\top(1)}_{0kj}\nabla^2\varphi^{,\,k}
+\frac{2\tau_0^2}{3}\varphi^{,\,kl}\chi^{\top(1)}_{0kj,\,l}
\nn\\
&
-\frac{\tau_0^2}{3}\varphi^{,\,kl}\chi^{\top(1)}_{0kl,j}
+\frac{2\tau_0}{3}\varphi^{,\,k}\chi^{\top(1)' }_{0kj}
-\frac{4}{3}\varphi^{,\,k}\chi^{\top(1)}_{0k j}
\nn\\
&
+\nabla^{-2}\partial_j\bigg(
\frac{2\tau_0^2}{3}\varphi^{,\,kl}\nabla^2\chi^{\top(1)}_{0kl}
+\frac{\tau_0^2}{3}\chi^{\top(1)}_{0kl,m}\varphi^{,\,klm}
-\frac{2\tau_0}{3}\varphi^{,\,kl}\chi^{\top(1)' }_{0kl}
+\frac{4}{3}\varphi^{,\,kl}\chi^{\top(1)}_{0kl}
\bigg)
\nn\\
&
-\frac{6}{\tau_0^{3}}\chi^{\top(1) }_{0kl,  \, j}\nabla^{-2}X^{,kl}
+\partial_j\nabla^{-2}
\bigg(
\frac{6}{\tau_0^{3}}\nabla^2\chi^{\top(1) }_{0kl}\nabla^{-2}X^{,kl}
+\frac{6}{\tau_0^{3}}\chi^{\top(1) }_{0kl, m}\nabla^{-2}X^{,klm}
\bigg) \,,
\el
 depending  on the initial values at $\tau_0$.
To get $ \chi^{\perp (2) }_{s(t)\,ij}$ from Eq.(\ref{chiSTperp}),
one has to remove $\partial^j$
as  follows.

Writing  $\chi^{\perp(2)}_{s(t)ij} = A_{s(t)i,j} + A_{s(t)j,i}$
in terms of a 3-vector $ A_{s(t)i }$
as  Eq.(\ref{chiVec1}),
 Eq.(\ref{chiSTperp})  yields
\bl
A_{{s(t)} j}
=&
\nabla^{-2}Q_j
+\nabla^{-2}\int_{\tau_0}^{\tau}\bigg[
-\frac{2\tau'}{3}\partial_j(\varphi^{,\,kl}\chi^{\top(1)}_{kl})
+\frac{2\tau'}{3}\nabla^2(\varphi^{,\,k}\chi^{\top(1)}_{kj})
\bigg]d\tau'
\nn\\
&
+\nabla^{-2}\bigg[
\frac{\tau^2}{3}\partial_j(\varphi^{,\,kl}\chi^{\top(1)}_{kl})
-\frac{\tau^2}{3}\chi^{\top(1)}_{kj}\nabla^2\varphi^{,\,k}
-\frac{2\tau^2}{3}\varphi^{,\,kl}\chi^{\top(1)}_{kj,\,l}
+\frac{\tau^2}{3}\varphi^{,\,kl}\chi^{\top(1)}_{kl,j}
\nn\\
&
-\frac{2\tau}{3}\varphi^{,\,k}\chi^{\top(1)' }_{kj}
+\frac{4}{3}\varphi^{,\,k}\chi^{\top(1)}_{k j}
\bigg]
+\nabla^{-2}\nabla^{-2}\partial_j\bigg[
-\frac{2\tau^2}{3}\varphi^{,\,kl}\nabla^2\chi^{\top(1)}_{kl}
\nn\\
&
-\frac{\tau^2}{3}\chi^{\top(1)}_{kl,m}\varphi^{,\,klm}
+\frac{2\tau}{3}\varphi^{,\,kl}\chi^{\top(1)' }_{kl}
-\frac{4}{3}\varphi^{,\,kl}\chi^{\top(1)}_{kl}
\bigg]
\nn\\
&
+\nabla^{-2} \int_{\tau_0}^\tau
\bigg[
-\frac{12}{\tau^{'3}}\chi^{\top(1)'}_{jk,\,l}\nabla^{-2}X^{,kl}
-\frac{6}{\tau^{'3}}\chi^{\top (1)' }_{kj}X^{,k}
\nn\\
&
+\partial_j\nabla^{-2}
\Big(
\frac{12}{\tau^{'3}}\chi^{\top(1)'}_{kl,m}\nabla^{-2}X^{,klm}
+\frac{6}{\tau^{'3}}\chi^{\top (1)' }_{kl}X^{,kl}
\Big)
\bigg] d\tau'
+\nabla^{-2}\bigg[
\frac{6}{\tau^{3}}\chi^{\top(1) }_{kl,  \, j}\nabla^{-2}X^{,kl}
\nn\\
&
-\partial_j\nabla^{-2}
\bigg(
\frac{6}{\tau^{3}}\nabla^2\chi^{\top(1) }_{kl}\nabla^{-2}X^{,kl}
+\frac{6}{\tau^{3}}\chi^{\top(1) }_{kl, m}\nabla^{-2}X^{,klm}
\bigg) \bigg]  \ .
\el
Thus, the vector perturbation  is obtained
\bl\label{chiSTperp2}
\chi^{\perp(2)}_{s(t)ij}
=&
\nabla^{-2} Q_{i,j}
+\int_{\tau_0}^{\tau} \frac{2\tau'}{3}\bigg[
 \partial_i(\varphi^{,\,k}\chi^{\top(1)}_{kj})
-\nabla^{-2}\partial_i\partial_j(\varphi^{,\,kl}\chi^{\top(1)}_{kl})
\bigg]d\tau'   \nn \\
&
+\nabla^{-2}\bigg[
\frac{ \tau^2}{3}\partial_i\partial_j(\varphi^{,\,kl}\chi^{\top(1)}_{kl})
+\partial_i\big( \frac{\tau^2}{3}\varphi^{,\,kl}\chi^{\top(1)}_{kl,j}
-\frac{\tau^2}{3}\chi^{\top(1)}_{kj}\nabla^2\varphi^{,\,k}
\nn  \\
&
-\frac{2\tau^2}{3}\varphi^{,\,kl}\chi^{\top(1)}_{kj,\,l}
-\frac{2\tau}{3}\varphi^{,\,k}\chi^{\top(1)' }_{kj}
+\frac{4}{3}\varphi^{,\,k}\chi^{\top(1)}_{k j}
\big)
\bigg]
+\nabla^{-2}\nabla^{-2}\partial_i\partial_j\bigg[
\frac{2\tau}{3}\varphi^{,\,kl}\chi^{\top(1)' }_{kl}
\nn \\
&
-\frac{2\tau^2}{3}\varphi^{,\,kl}\nabla^2\chi^{\top(1)}_{kl}
-\frac{\tau^2}{3}\chi^{\top(1)}_{kl,m}\varphi^{,\,klm}
-\frac{4}{3}\varphi^{,\,kl}\chi^{\top(1)}_{kl}
\bigg]
\nn\\
&
+\nabla^{-2} \int_{\tau_0}^\tau
\bigg[
\partial_j
\Big(
-\frac{12}{\tau^{'3}}\chi^{\top(1)'}_{ki,\,l}\nabla^{-2}X^{,kl}
-\frac{6}{\tau^{'3}}\chi^{\top (1)' }_{ki}X^{,k}
\Big)
\nn\\
&
+\partial_i\partial_j\nabla^{-2}
\Big(
\frac{12}{\tau^{'3}}\chi^{\top(1)'}_{kl,m}\nabla^{-2}X^{,klm}
+\frac{6}{\tau^{'3}}\chi^{\top (1)' }_{kl}X^{,kl}
\Big)
\bigg]d\tau'
\nn\\
&
+\nabla^{-2}\bigg[
\partial_j
\Big(
\frac{6}{\tau^{3}}\chi^{\top(1) }_{kl,  \, i}\nabla^{-2}X^{,kl}
\Big)
-\partial_i\partial_j\nabla^{-2}
\bigg(
\frac{6}{\tau^{3}}\nabla^2\chi^{\top(1) }_{kl}\nabla^{-2}X^{,kl}
\nn\\
&
+\frac{6}{\tau^{3}}\chi^{\top(1) }_{kl, m}\nabla^{-2}X^{,klm}
 \bigg) \bigg]
+ (i \leftrightarrow j) \,  .
\el
Actually,
this  vector mode   $\chi^{\perp(2) }_{s(t)ij} $
can be also derived from  the curl portion of
the momentum constraint (\ref{cfta1q}) itself
without
explicitly using the solutions
$\phi^{(2)}_{s(t)}$ of (\ref{phi2st})
and $\chi^{\parallel(2)}_{s(t)}$ of  (\ref{chi2st||1}).
The result (\ref{chiSTperp2}) explicitly demonstrates that
the 2nd-order vector perturbation exists,
whose  effective source is  the coupling of 1st-order perturbations,
even though  the   matter source  for the vector mode is zero
in the synchronous gage,
$T_{0i}=  0$, $T_{ij}=  0$.

\subsection{ Tensor Perturbation   $\chi^{\top(2)}_{s(t)i j}$  }

Next consider the traceless part of the evolution equation (\ref{afk1q})
\bl\label{RGWst1}
\chi^{\top (2)''}_{s(t)ij}
    +\frac{4}{\tau}\chi^{\top (2)'}_{s(t)ij}
    -\nabla^2\chi^{\top(2)}_{s(t)i j}
=&2\bar S_{s(t)ij}
-\l(2D_{ij}\phi^{(2)}_{s(t)}
+\frac{1}{3}\nabla^2D_{ij}\chi^{||(2)}_{s(t)}
\r)
\nonumber \\
&
-\l( D_{ij}\chi^{||(2)''}_{s(t)}
 +\frac{4}{\tau}D_{ij}\chi^{||(2)'}_{s(t)}\r)
-\l(\chi^{\perp (2)''}_{s(t)ij}
    +\frac{4}{\tau}\chi^{\perp (2)'}_{s(t)ij}
\r)
,
\el
where $\bar{S}_{s(t)ij}$ is the traceless part of
$ S_{s(t)ij}  $ as the following
\bl\label{barSstij}
\bar{S}_{s(t)ij}=&
-\frac{2\tau}{3} \varphi^{,k}_{,i}\chi^{\top(1)'}_{kj}
-\frac{2\tau}{3} \varphi_{,j}^{,k}\chi^{\top(1)'}_{k i}
+\frac{4\tau}{9} \varphi^{,kl}\chi^{\top(1)'}_{kl}\delta_{ij}
+\frac{\tau}{3} \chi^{\top(1)'}_{ij} \nabla^2 \varphi
+\frac{10}{3}\chi^{\top(1)}_{ij}  \nabla^2 \varphi
\nn\\
&
+\frac{20}{9}\varphi^{,kl}\chi^{\top(1)}_{kl}\delta_{ij}
-\frac{10}{3}\varphi_{,j}^{,k}\chi^{\top(1)}_{ki}
-\frac{10}{3}\varphi_{,i}^{,k}\chi^{\top(1)}_{kj}
+\frac{10}{3}  \varphi\nabla^2\chi^{\top(1)}_{ij}
+\frac{\tau^2}{3} \varphi^{,kl}\chi^{\top(1)}_{ij,\,kl}
\nn\\
&
+\frac{\tau^2}{3} \varphi^{,kl}\chi^{\top(1)}_{kl,\,ij}
-\frac{\tau^2}{9} \varphi^{,kl}\nabla^2\chi^{\top(1)}_{kl}\delta_{ij}
-\frac{\tau^2}{3} \varphi^{,kl}\chi^{\top(1)}_{li,\,jk}
-\frac{\tau^2}{3} \varphi^{,kl}\chi^{\top(1)}_{lj,\,ik}
+5  \varphi^{,\,k}\chi^{\top(1)}_{ij,\,k}
\nn
\\
&
-\frac{5}{3}  \varphi^{,\,k}\chi^{\top(1)}_{ki,j}
-\frac{5}{3}  \varphi^{,\,k}\chi^{\top(1)}_{kj,\,i}
+\frac{\tau^2}{6}  \chi^{\top(1)}_{ij,\,k}\nabla^2 \varphi^{,\,k}
-\frac{\tau^2}{6}\chi^{\top(1)}_{ki,\,j} \nabla^2\varphi^{,k}
- \frac{\tau^2}{6}  \chi^{\top(1)}_{kj,\,i}\nabla^2 \varphi^{,\,k}
\nn\\
&
+\frac{\tau^2}{6} \varphi^{,kl}_{,\,i}\chi^{\top(1)}_{kl,\,j}
+\frac{\tau^2}{6} \varphi^{,kl}_{,j}\chi^{\top(1)}_{kl,\,i}
-\frac{\tau^2}{9} \varphi^{,klm}\chi^{\top(1)}_{kl,\,m}\delta_{ij}
\nn\\
&
-\frac{12}{\tau^4}\chi^{\top(1)'}_{kl}\nabla^{-2}X^{,kl}\delta_{ij}
-\frac{2}{\tau^3}\chi^{\top(1)}_{kl,\,m}\nabla^{-2}X^{,klm}\delta_{ij}
-\frac{2}{\tau^3}\nabla^2\chi^{\top(1)}_{kl}\nabla^{-2}X^{,kl}\delta_{ij}
\nn\\
&
-\frac{9}{\tau^4}X\chi^{\top(1)'}_{ij}
+\frac{18}{\tau^4}\chi^{\top(1)'}_{k i}\nabla^{-2}X^{,k}_{,j}
+\frac{18}{\tau^4}\chi^{\top(1)'}_{kj}\nabla^{-2}X^{,k}_{,i}
\nn\\
&
-\frac{3}{\tau^3}X^{,\,k}\chi^{\top(1)}_{k j,\,i}
-\frac{3}{\tau^3}X^{,\,k}\chi^{\top(1)}_{k i,\,j}
+\frac{3}{\tau^3}X^{,\,k}\chi^{\top(1)}_{ij,\,k}
\nn\\
&
-\frac{6}{\tau^3}\chi^{\top(1)}_{lj,\,ik}\nabla^{-2}X^{,kl}
-\frac{6}{\tau^3}\chi^{\top(1)}_{li,\,jk}\nabla^{-2}X^{,kl}
+\frac{6}{\tau^3}\chi^{\top(1)}_{ij,\,kl}\nabla^{-2}X^{,kl}
\nn\\
&
+\frac{6}{\tau^3}\chi^{\top(1)}_{kl,\,ij}\nabla^{-2}X^{,kl}
+\frac{3}{\tau^3}\chi^{\top(1)}_{kl,\,i}\nabla^{-2}X^{,kl}_{,j}
+\frac{3}{\tau^3}\chi^{\top(1)}_{kl,\,j}\nabla^{-2}X^{,kl}_{,\,i}
\  .
\el
One can substitute the known  $\phi^{(2)}_{s(t)}$,
$ D_{ij}\chi^{||(2) }_{s(t)} $,
$ \chi^{\perp (2)}_{s(t)ij}$ into  Eq.(\ref{RGWst1}),
and solve for $\chi^{\top (2)}_{s(t)ij}$.
But the following calculation is   simpler and will yield the same result.
Applying $\partial^i\partial^j$ to (\ref {RGWst1}) yields
\be\label{RGWst2}
-\l(2D_{ij}\phi^{(2)}_{s(t)}
+\frac{1}{3}\nabla^2D_{ij}\chi^{||(2)}_{s(t)}
\r)
- \l(D_{ij}\chi^{||(2)''}_{s(t)}
 +\frac{4}{\tau}D_{ij}\chi^{||(2)'}_{s(t)}\r)
=-3D_{ij}\nabla^{-2}\nabla^{-2}\bar S_{s(t)kl}^{,\,kl}
\,.
\ee
Substituting Eq.(\ref{RGWst2}) into the rhs of Eq.(\ref{RGWst1}),
one has
\be\label{RGWst3}
\chi^{\top (2)''}_{s(t)ij}
    +\frac{4}{\tau}\chi^{\top (2)'}_{s(t)ij}
    -\nabla^2\chi^{\top(2)}_{s(t)i j}
=2\bar S_{s(t)ij}
-3D_{ij}\nabla^{-2}\nabla^{-2}\bar S_{s(t)kl}^{,\,kl}
-\l(\chi^{\perp (2)''}_{s(t)ij}
    +\frac{4}{\tau}\chi^{\perp (2)'}_{s(t)ij}
\r)
.
\ee
Applying $\partial^j$ to (\ref {RGWst3}) and
together with Eq.(\ref{chiVec1})
leads  to an equation of  $A_{s(t)i}$:
\be\label{Aperp2}
0=2\bar S_{s(t)ij}^{,j}
-2\nabla^{-2}\bar S_{s(t)kl,\,i}^{,\,kl}
-\nabla^2\l(A^{''}_{s(t)i}
    +\frac{4}{\tau}A^{'}_{s(t)i}
\r)
.
\ee
Thus,
from Eq.(\ref{chiVec1}) and Eq.(\ref{Aperp2}),
one has
\bl\label{RGWst4}
-\l(\chi^{\perp (2)''}_{s(t)ij}
    +\frac{4}{\tau}\chi^{\perp (2)'}_{s(t)ij}
\r)
=&-\partial_j\l(A^{''}_{s(t)i}
    +\frac{4}{\tau}A^{'}_{s(t)i}\r)
-\partial_i\l(A^{''}_{s(t)j}
    +\frac{4}{\tau}A^{'}_{s(t)j}\r)
\nn\\
=&-2\nabla^{-2}\bar S_{s(t)ki,j}^{,k}
-2\nabla^{-2}\bar S_{s(t)kj,\,i}^{,k}
+4\nabla^{-2}\nabla^{-2}\bar S_{s(t)kl,\,ij}^{,\,kl}
\ .
\el
Substituting (\ref{RGWst4})
into the rhs of Eq.(\ref{RGWst3}),
we obtain the equation for $\chi^{\top (2)}_{s(t)ij}$ :
\bl\label{gw11}
\chi^{\top (2)''}_{s(t)ij}
    +\frac{4}{\tau}\chi^{\top (2)'}_{s(t)ij}
    -\nabla^2\chi^{\top(2)}_{s(t)i j}
=&J_{s(t)ij}({\bf x},\tau)
\el
with the source
\be
   J_{s(t)ij}({\bf x},\tau)
 \equiv   2\bar S_{s(t)ij}
+\nabla^{-2}\nabla^{-2}\bar S_{s(t)kl,\,ij}^{,\,kl}
+\delta_{ij}\nabla^{-2}\bar S_{s(t)kl}^{,\,kl}
-2\nabla^{-2}\bar S_{s(t)ki,j}^{,k}
-2\nabla^{-2}\bar S_{s(t)kj,\,i}^{,k}
\ .
\ee
where the known symmetric and traceless
$\bar S_{s(t)ij}$ is given by (\ref{barSstij}).
It is checked   that $J_{s(t)ij}({\bf x},\tau)$ is
traceless and transverse.

The differential equation (\ref{gw11}) is inhomogeneous,
and its solution is given by
\be \label{solgw}
\chi^{\top(2)}_{s(t)i j}  ({\bf x},\tau)
 =  \frac{1}{(2\pi)^{3/2} }
   \int d^3k   e^{i \,\bf{k}\cdot\bf{x}}
      \left(\bar I_{s(t)ij} (s)
           +  \frac{b_{1ij}}{ s^{ 3/2}}   H^{(1)}_{\frac{3}{2}}(s)
           +  \frac{b_{2ij}}{ s^ {3/2}}   H^{(2)}_{\frac{3}{2}}(s)
           \right),
\ee
where  $s \equiv k\tau$,
\bl
\bar I_{s(t)ij}( s)
 \equiv &  \frac{1}{s^2}
  (\cos s-\frac{\sin s}{ s} )
            \int^s_1 dy \, y^2 ( \sin y + \frac{\cos y}{y} ) \bar J_{s(t) ij} (y) \nn \\
& - \frac{1}{s^2} ( \sin s + \frac{\cos s}{s})
            \int^s_1 dy \,  y^2 ( \cos y - \frac{\sin y}{y} ) \bar J_{s(t) ij}(y),
\el
with $\bar J_{s(t)ij}$ being the Fourier transformation of the source $ J_{s(t)ij} $.
In (\ref{solgw})
the two terms associated with   $b_{1ij}$ and  $b_{2ij}$
are of  the same form as the 1st-order solution
$\chi^{\top(1)}_{ij}  ({\bf x},\tau)$
in (\ref{Fourier}) and  (\ref{GWmode})
and correspond to the homogeneous solution of (\ref{gw11}).
These  two terms are kept in order to allow for a general
 initial condition at time $\tau_0$.
In particular,  the coefficients $b_{1ij}$ and  $b_{2ij}$
are to be determined by the 2nd-order tensor modes of precedent Radiation Dominated stage.

Thus,  all the 2nd-order metric perturbations
due to scalar-tensor coupling  have been obtained.
By (\ref{density 2order}),
the corresponding 2nd-order density contrast is
\be
\delta^{(2)}_{s(t)}=
\delta^{(2)}_{s(t)0}
+3(\phi^{(2)}_{s(t)}-\phi^{(2)}_{s(t)0})
+ (\chi^{\top(1)\, ij}D_{ij}\chi^{\parallel(1)}
              -  \chi^{\top(1)\, ij}_0  D_{ij}\chi^{\parallel(1)}_0),
\ee
which can be expressed as
\bl\label{deltaST2unEx2}
\delta^{(2)}_{s(t)}
    =&
-\frac{\tau^2}{20}C
-\frac{2}{3\tau^3}\varphi^{,kl}\int^{\tau}_{\tau_0}
    \tau^{'4}\,\chi^{\top(1) }_{kl} d\tau'
-\frac{9}{20\tau^3}(\nabla^{-2}X^{,kl})
    \int^\tau_{\tau_0}\tau'\nabla^2\chi^{\top(1)}_{kl}d\tau'
\nn\\
&
 +\frac{3}{\tau^3}W
-\frac{9}{4\tau^3}\chi^{\top(1)}_{kl}\nabla^{-2}X^{,kl}
+\frac{9\tau^2}{20}(\nabla^{-2}X^{,kl})
    \int_{\tau_0}^{\tau}
    \frac{1}{\tau^{'4}}\nabla^2\chi^{\top(1)}_{kl}d\tau'
           \ .
\el
 after using the given $\phi^{(2)}_{s(t)} $,
 $ D_{ij}\chi^{\parallel(1)}$,  $\chi^{\top(1)\, ij}$.

\section{ 2nd-Order   Perturbations
             with the Source  $\chi^{\top(1)}_{kl}\chi^{\top(1)}_{ij}$}

Now we turn to  the set of Eqs.(\ref{ens26})--(\ref{eq34s26})
 with  the source of the form of $\chi^{\top(1)}_{kl}\chi^{\top(1)}_{ij}$,
and  derive
the solution of the   second-order perturbations.
The procedures involved are  similar to those in Sec. 4.

\subsection{  Scalar Perturbation  $\phi^{(2)}_{T}$   }

Combing  the constraint equations,
 i.e,
 (\ref{ens26}) + $\frac{1}{6}\partial^j \int_{\tau_0} ^\tau d\tau $ (\ref{mos26}),
using the 1st-order GW equation (\ref{eqRGW1}),
one has the following  differential equation of $ \phi^{(2)}_T$:
\bl\label{phi2TEq}
\phi^{(2)'}_{T}
+\frac{3}{\tau}\phi^{(2)}_{T}
          =&
-\frac{1}{3}\chi^{\top(1) ' ij}\chi^{\top(1)}_{ij}
 -\frac{1}{2\tau}\chi^{\top(1) ij}\chi^{\top(1)}_{ij}
+\frac{\tau}{6}\int^{\tau}_{\tau_0}
    \frac{\chi^{\top(1)'}_{kl}\chi^{\top(1)'kl}}{\tau'} d\tau'
\nonumber  \\
 &+\frac{1}{\tau}\bigg(3\phi^{(2)}_{T\,0}
-\delta_{T\,0}^{(2)}
 +\frac{1}{2}\chi^{\top(1) ij}_0 \chi^{\top(1)}_{0ij}
\bigg)
-\frac{\tau}{12} K
\,,
\el
where the constant
\bl\label{KT}
K\equiv&
-2\nabla^2\phi^{(2)}_{{T}0}
-\frac{1}{3}\nabla^2\nabla^2\chi^{\parallel(2)\, }_{{T}0}
+\frac{1}{2}\chi^{\top(1)kl,m}_0\chi^{\top(1)}_{0km,\,l}
-\frac{3}{4}\chi^{\top(1)kl,m}_0\chi^{\top(1)}_{0kl,m}
\nn\\
&
-\chi^{\top(1)kl}_0\nabla^2\chi^{\top(1)}_{0kl}
+\frac{1}{4}\chi^{\top(1)'}_{0\, kl}\chi_0 ^{\top(1)'kl} \ ,
\el
depends on the initial metric perturbations at $\tau_0$.
The solution of Eq.(\ref{phi2TEq}) is
\bl\label{phi2T}
\phi^{(2)}_{T}
=&
\bigg(
\phi^{(2)}_{T\,0}
-\frac{1}{3}\delta_{T\,0}^{(2)}
 +\frac{1}{6}\chi^{\top(1) ij}_0 \chi^{\top(1)}_{0ij}
 \bigg)
-\frac{\tau^2}{60}K   +\frac{B({\bf x})}{\tau^3}
-\frac{1}{6}\chi^{\top(1)  ij}\chi^{\top(1)}_{ij} \nn \\
&
+\frac{\tau^{2}}{30}\int_{\tau_0}^{\tau}
    \frac{\chi^{\top(1)'}_{kl}\chi^{\top(1)'kl}}{\tau'}  d\tau'
-\frac{1}{30\tau^3}\int^{\tau}_{\tau_0}\tau^{'4}
          \chi^{\top(1)'}_{kl}\chi^{\top(1)'kl}d\tau'
 \,     ,
\el
where  integration by parts has been  used,
and $B({\bf x})$ is fixed by setting
 $\tau=\tau_0$ and $\phi^{(2)}_{T}(\tau_0)=\phi^{(2)}_{T0}$ as
\be\label{BdecayMode0}
B({\bf x})  =\frac{\tau_0^3}{3}\delta_{T\,0}^{(2)}
            +\frac{\tau_0^5}{60}K \, .
\ee
Notice that the solution (\ref{phi2T}) can be also derived by
 the trace part of the evolution equation (\ref{eq34s26})
  together with
the energy constraint (\ref{ens26}).

\subsection{  Scalar Perturbation $\chi^{||(2)}_{T}$}

The expression  $  \partial^j \int_{\tau_0} ^\tau d\tau' $ Eq.(\ref{mos26}) gives
the following equation
\bl\label{mosIntTT}
2\nabla^2\phi^{(2)}_{T}
 +\frac{1}{2}   D_{ij}\chi^{\parallel(2),\, ij }_{T}
 =&\int_{\tau_0}^\tau d\tau' M_{Tj}^{,j}
+2\nabla^2\phi^{(2)}_{T0}
+\frac{1}{3}\nabla^2\nabla^2\chi^{\parallel(2)\, }_{T0} ,
\el
Substituting $M_{T j}$ of (\ref{MTj}) and
$\phi^{(2)}_{T}$ of Eq.(\ref{phi2T})
into Eq.(\ref{mosIntTT}) yields
\bl\label{chi2st||1TT}
\chi^{\parallel(2)\, }_{T}
      =& Y +\frac{\tau^2}{10}\nabla^{-2}K -\frac{6}{\tau^3}\nabla^{-2}B
+\nabla^{-2}\bigg[
-\frac{1}{2}\chi^{\top(1)  kl}\chi^{\top(1)}_{kl}
-\frac{\tau^{2}}{5}\int_{\tau_0}^{\tau}
    \frac{\chi^{\top(1)'}_{kl}\chi^{\top(1)'kl}}{\tau'} d\tau'
\nn\\
&
+\frac{1}{5\tau^3}\int ^{\tau}_{\tau_0}\tau^{'4}\chi^{\top(1)'}_{kl}\chi^{\top(1)'kl}d\tau'
\bigg]
+\nabla^{-2}\nabla^{-2}\bigg[
\frac{3}{2}\chi^{\top(1)kl,m}\chi^{\top(1)}_{km,\,l}  \nn \\
&
+\frac{3}{4}\chi^{\top(1)kl,m}\chi^{\top(1)}_{kl,m}
+\frac{3}{4}\chi^{\top(1)'}_{kl}\chi^{\top(1)'kl}
+6\int_{\tau_0}^{\tau}\frac{1}{\tau'}\chi^{\top(1)'}_{kl}\chi^{\top(1)'kl}d\tau'
\bigg]
,
\el
where the constant
\bl
Y \equiv  &
\chi^{\parallel(2)\, }_{{T}0}
+\nabla^{-2}\bigg(
2\delta_{T\,0}^{(2)}
-\chi^{\top(1) ij}_0 \chi^{\top(1)}_{0ij}
 \bigg)
+\nabla^{-2}\nabla^{-2}\bigg(
-\frac{3}{2}\chi^{\top(1)kl,m}_0\chi^{\top(1)}_{0km,\,l}
\nn\\
&
+\frac{9}{4}\chi^{\top(1)kl,m}_0\chi^{\top(1)}_{0kl,m}
+3\chi^{\top(1)kl}_0\nabla^2\chi^{\top(1)}_{0kl}
-\frac{3}{4}\chi^{\top(1)'}_{0\, kl}\chi_0^{\top(1)'kl}
\bigg) ,
\el
depending on the initial values of metric perturbations at $\tau_0$.
Thus,
 the scalar perturbation   $D_{ij}\chi^{||(2) }_T$ is   determined,
\bl\label{chi112qqqTT}
D_{ij}\chi^{\parallel(2)\, }_{T}
      =& D_{ij}Y
+\frac{\tau^2}{10}D_{ij}\nabla^{-2}K
-\frac{6}{\tau^3}D_{ij}\nabla^{-2}B
  \nn \\
&
+D_{ij}\nabla^{-2}\bigg[
-\frac{1}{2}\chi^{\top(1)  kl}\chi^{\top(1)}_{kl}
-\frac{\tau^{2}}{5}\int_{\tau_0}^{\tau}
    \frac{1}{\tau'}\chi^{\top(1)'}_{kl}\chi^{\top(1)'kl} d\tau'
\nn\\
&
+\frac{1}{5\tau^3}\int ^{\tau}_{\tau_0}\tau^{'4}\chi^{\top(1)'}_{kl}\chi^{\top(1)'kl} d\tau'
\bigg]
+D_{ij}\nabla^{-2}\nabla^{-2}\bigg[
\frac{3}{2}\chi^{\top(1)kl,m}\chi^{\top(1)}_{km,\,l} \nn \\
&
+\frac{3}{4}\chi^{\top(1)kl,m}\chi^{\top(1)}_{kl,m}
+\frac{3}{4}\chi^{\top(1)'}_{kl}\chi^{\top(1)'kl}
+6\int_{\tau_0}^{\tau}\frac{1}{\tau'}\chi^{\top(1)'}_{kl}\chi^{\top(1)'kl} d\tau'
\bigg] \,  .
\el
We have checked that when $B({\bf x})$ satisfies
Eq.(\ref{BdecayMode0}),
at the initial time $\tau=\tau_0$,
one has
$\chi^{||(2) }_{T}=\chi^{||(2) }_{T0} $.
Notice that the solution (\ref{chi112qqqTT}) can also be derived by
the traceless part of the evolution equation (\ref{eq34s26}) together with
the momentum constraint (\ref{mos26}).

\subsection{Vector Perturbation  $\chi^{\perp(2)}_{Ti j}$  }

The time integral of the momentum constraint (\ref{mos26})
from $\tau_0$ to $\tau$ is
\be\label{mosInt2TT}
2\phi^{(2)}_{T,j}
     +\frac{1}{3}\nabla^2\chi^{\parallel(2)}_{T,j}
+\frac12\chi^{\perp (2),\,i}_{{T}i j}
               =\int_{\tau_0}^{\tau}d\tau' M_{Tj}
               +2\phi^{(2)}_{T0,j}
     +\frac{1}{3}\nabla^2\chi^{\parallel(2)}_{T0,j}
+\frac12\chi^{\perp (2),\,i}_{{T}0i j}
\ ,
\ee
Using $M_{Tj}$ in (\ref{MTj}),
one has
\be
\int_{\tau_0}^\tau d\tau' M_{Tj}
=\frac{1}{2}\int_{\tau_0}^{\tau}P_{j}(\mathbf x,\tau')d\tau'
-\chi^{\top(1) kl}\chi^{\top(1)}_{kl,\,j}
+\chi^{\top(1) kl}_0\chi^{\top(1)}_{0kl,\,j}
,
\ee
where
\be
P_{j}(\mathbf x,\tau)\equiv
2\chi^{\top(1) kl}\chi^{\top(1)'}_{j k,\,l}
+\chi^{\top(1) kl}_{,\,j}\chi^{\top(1)'}_{kl}  .
\ee
Plugging the solutions   $\phi^{(2)}_{T}$ of  (\ref{phi2T})
and $\chi^{\parallel(2)}_{T}$ of  (\ref{chi112qqqTT})
into (\ref{mosInt2TT}),
after calculations similar to Sec. 4.3,
  the vector perturbation  is obtained:
\bl\label{chiTTperp2}
\chi^{\perp(2)}_{Tij}
=&
\nabla^{-2}(N_{i,j}+N_{j,i})
+\nabla^{-2}\int_{\tau_0}^{\tau}
[ \partial_i P_{j} +\partial_j P_{i} ]d\tau'
\nn\\
&
-\partial_i\partial_j\nabla^{-2}\nabla^{-2}\bigg[
2\chi^{\top(1)kl,m}\chi^{\top(1)}_{km,\,l}
+\chi^{\top(1)kl,m}\chi^{\top(1)}_{kl,m}
\nn\\
&
+\chi^{\top(1)'}_{kl}\chi^{\top(1)'kl}
+8\int_{\tau_0}^{\tau}\frac{\chi^{\top(1)'}_{kl}\chi^{\top(1)'kl}}{\tau'} d\tau'
\bigg]
\ ,
\el
where the constant 3-vector
\be
N_j\equiv\chi^{\perp (2),\,i}_{{T}0i j}
+\partial_j\nabla^{-2}\bigg(
\chi^{\top(1)kl,m}_0\chi^{\top(1)}_{0km,\,l}
+\frac{1}{2}\chi^{\top(1)kl,m}_0\chi^{\top(1)}_{0kl,m}
+\frac{1}{2} \chi^{\top(1)'}_{0\, kl}\chi_0 ^{\top(1)'kl}
\bigg)
\ ,
\ee
  depending  on the initial values at $\tau_0$.
Notice that the solution (\ref{chiTTperp2}) can be also derived by
 the traceless part of the evolution equation (\ref{eq34s26})
 together with
the momentum constraint (\ref{mos26}).

\subsection{ Tensor Perturbation   $\chi^{\top(2)}_{Ti j}$  }

Next consider the traceless part of the evolution equation  (\ref{eq34s26})
\bl\label{RGWTT1}
\chi^{\top (2)''}_{Tij}
    +\frac{4}{\tau}\chi^{\top (2)'}_{Tij}
    -\nabla^2\chi^{\top(2)}_{Ti j}
=&2\bar S_{Tij}
- (2D_{ij}\phi^{(2)}_{T}
+\frac{1}{3}\nabla^2D_{ij}\chi^{||(2)}_{T}   )
\nonumber \\
&
-( D_{ij}\chi^{||(2)''}_{T}
 +\frac{4}{\tau}D_{ij}\chi^{||(2)'}_{T}  )
- (\chi^{\perp (2)''}_{Tij}  +\frac{4}{\tau}\chi^{\perp (2)'}_{Tij}  ),
\el
where
\bl\label{barSTTij}
\bar{S}_{Tij} \equiv  & S_{Tij}-\frac{1}{3}\delta_{ij}S^k_{Tk} \nn \\
= &
\chi^{\top(1)'k}_{i}\chi^{\top(1)'}_{kj}
-\frac13\chi^{\top(1)'kl}\chi^{\top(1)'}_{kl}\delta_{ij}
+\chi^{\top(1)kl}\chi^{\top(1)}_{li,\,jk}
+\chi^{\top(1)kl}\chi^{\top(1)}_{lj,\,ik}
\nn\\
&
-\chi^{\top(1)kl}\chi^{\top(1)}_{kl,\,ij}
-\chi^{\top(1)kl}\chi^{\top(1)}_{ij,\,kl}
+\chi^{\top(1),\,k}_{l i}\chi^{\top(1),\,l}_{k j}
-\chi^{\top(1)k}_{i,\,l}\chi^{\top(1),\,l}_{j k}
\nn\\
&
-\frac12\chi^{\top(1)kl}_{,\,i}\chi^{\top(1)}_{kl,\,j}
+\frac12\chi^{\top(1)kl,\,m}\chi^{\top(1)}_{kl,\,m}\delta_{ij}
-\frac13\chi^{\top(1)}_{ml,\,k}\chi^{\top(1)m k,\,l}\delta_{ij}
    \nn\\
&
+\frac13\chi^{\top(1)kl}\nabla^2\chi^{\top(1)}_{kl}\delta_{ij}
.
\el
By calculations similar to   Sec. 4.4,
 Eq.(\ref{RGWTT1}) is written as
\bl\label{gw11TT}
\chi^{\top (2)''}_{Tij}
    +\frac{4}{\tau}\chi^{\top (2)'}_{Tij}
    -\nabla^2\chi^{\top(2)}_{Ti j}
=&J_{Tij}({\bf x},\tau)
\el
where the source
\be
J_{Tij}({\bf x},\tau)
   \equiv   2\bar S_{Tij}
+\nabla^{-2}\nabla^{-2}\bar S_{Tkl,\,ij}^{,\,kl}
+\delta_{ij}\nabla^{-2}\bar S_{Tkl}^{,\,kl}
-2\nabla^{-2}\bar S_{Tki,j}^{,k}
-2\nabla^{-2}\bar S_{Tkj,\,i}^{,k}
\ .
\ee
The differential equation (\ref{gw11TT}) is inhomogeneous,
and its solution is given by
\be \label{solgwTT}
\chi^{\top(2)}_{Ti j}  ({\bf x},\tau)
 =  \frac{1}{(2\pi)^{3/2} }
   \int d^3k   e^{i \,\bf{k}\cdot\bf{x}}   \left( \bar I_{Tij} (s)
   + \frac{c_{1ij}}{ s^{ 3/2}}   H^{(1)}_{\frac{3}{2}}(s)
   +  \frac{c_{2ij}}{ s^ {3/2}}   H^{(2)}_{\frac{3}{2}}(s)       \right),
\ee
where  $s \equiv k\tau$,
\bl
\bar I_{Tij}( s)
 \equiv &  \frac{1}{s^2}
  (\cos s-\frac{\sin s}{s} )
            \int^s_1 dy \, y^2 ( \sin y + \frac{\cos y}{y} ) \bar J_{T ij} (y) \nn \\
& - \frac{1}{s^2} ( \sin s + \frac{\cos s}{s})
            \int^s_1 dy \,  y^2 ( \cos y - \frac{\sin y}{y} ) \bar J_{T ij}(y),
\el
with $\bar J_{Tij}$ being the Fourier transformation of the source $ J_{Tij} $.
In (\ref{solgwTT})  $c_{1ij}$ and $c_{2ij}$ terms
represent  a   homogeneous solution,
which should be determined by the initial condition at $\tau_0$.

Thus,  all the 2nd-order metric perturbations
produced by tensor-tensor coupling
have been obtained.
Consequently,  by (\ref{density 2order}),
the corresponding 2nd-order density contrast
\be
\delta^{(2)} =
 \delta^{(2)}_{T0}
+3 (\phi^{(2)}_{T} -\phi^{(2)}_{T0})
+\frac12
   ( \chi^{\top(1)\, ij} \chi^{\top(1)}_{ij}
   - \chi^{\top(1)\, ij}_0  \chi^{\top(1)}_{0\, ij}) .
\ee
which can be written as
\be
\delta^{(2)}_{T}
        = -\frac{\tau^2}{20}K  +\frac{3}{\tau^3}B
-\frac{1}{10\tau^3}\int^{\tau}_{\tau_0}\tau^{'4}\chi^{\top(1)'}_{kl}\chi^{\top(1)'kl}d\tau'
+\frac{\tau^{2}}{10}\int_{\tau_0}^{\tau}
    \frac{\chi^{\top(1)'}_{kl}\chi^{\top(1)'kl}}{\tau'}  d\tau'  .
\ee

So far,
the 2nd-order metric perturbations have been obtained
using the scalar-scalar, scalar-tensor, and tensor-tensor couplings.
We can qualitatively assess which coupling is dominant during MD stage.
By the solution (20) in the paper I \cite{ScalarScalar}
of the 1st order of perturbations,
the scalar is $\propto \tau^2 $, increasing with time,
the tensor  by (19) in the paper I \cite{ScalarScalar}
is
$\propto  \tau^{-\frac32 }H^{(1)}_{\frac{3}{2} }(k\tau ),
\tau^{-\frac32 } H^{(2)}_{\frac{3}{2} } (k\tau )$,
whose amplitude is decreasing with time.
So,
the scalar-scalar terms are increasing as $\tau^4$,
the tensor-tensor terms are decreasing quickly,
 and the tensor-scalar terms behave as
$\propto \tau^{1/2} H^{(1)}_{\frac{3}{2} }(k\tau ),
\tau^{1/2} H^{(2)}_{\frac{3}{2} } (k\tau )$,
which are decreasing over the whole range
at a slower rate than the tensor-tensor terms.
Thus, qualitatively speaking,
the scalar-scalar terms are dominant over the tensor terms  during evolution.
Therefore, the corresponding solutions of metric perturbations also share
these  generic features.

In  applications,
 one has to deal with the second-order degrees of gauge freedom in these  solutions,
 which is discussed in the latter section.

\section{The 2nd-Order  Gauge Transformations }

Consider  the coordinate transformation up to 2nd order \cite{Matarrese98,ScalarScalar}:
\be\label{xmutransf}
x^\mu \rightarrow  \bar x^\mu = x^\mu +\xi^{(1)\mu}
+ \frac{1}{2}\xi^{(1)\mu}_{,\alpha}\xi^{(1)\alpha}
+ \frac{1}{2}\xi ^{(2)\mu},
\ee
where $\xi^{(1)}$ is a  1st-order vector field,
      $\xi^{(2)}$ is a 2nd-order vector field,
  and   can be written in terms of
their respective parameters
\bl
&\xi^{(A)0}=\alpha^{(A)}\label{alpha_r},
\\
&\xi^{(A)i}=\partial^i\beta^{(A)}+d^{ (A)i} , \label{xi_r}
\el
with $A=1,2$ and a constraint $\partial_i d^{(A)i}=0$.
The 1st-order gauge transformations
between two synchronous coordinate systems
for the Einstein-de Sitter model with  $a(\tau)\propto \tau^2$
are listed in (C26)--(C30) in Ref.~\cite{ScalarScalar}.
The   general 2nd-order synchronous-to-synchronous gauge transformations
of metric perturbations are given by
  (C50) (C56)--(C58) of Ref.~\cite{ScalarScalar},
which are valid for general cosmic expansion stages.
In the following we apply them to the case of  $a(\tau)\propto \tau^2$.
So far in our paper,
the   perturbations of 4-velocity of dust
have been  taken to be   $U^{(1)\mu} = U^{(2)\mu} =0 $.
It is proper to require also
the transformed 3-velocity perturbations
\be \label{U10}
\bar U^{(1)i } =0,
\ee
\be  \label{U20}
\bar U^{(2) i } = 0 ,
\ee
in the new synchronous coordinate  \cite{Russ1996} \cite{ScalarScalar}.
Under the constraints  (\ref{U10}),
the 1st-order vector field  $\xi^{(1)\mu}$ is \cite{ScalarScalar}
\be\label{alpha1}
\alpha^{(1)}   = \frac{A^{(1)} }{\tau^2} \, ,
\ee
\be\label{beta1}
\beta^{(1)}_{,\,i}  = C^{||(1)} (\mathbf x)_{,\,i}
\ee
\be\label{d1}
d^{(1)}_i=0 \, .
\ee
In the above,  $A^{(1)} $ is an arbitrary constant,
and  $ C^{\parallel (1)}({\bf x})  $ is an arbitrary function.
The 1st-order residual gauge transformations are  \cite{ScalarScalar}
\ba \label{phigauge}
&& \bar \phi^{(1)}  =
   \phi^{(1)} +    2  \frac{ A^{(1)}}{\tau^3 }
   + \frac{1}{3}\nabla^2 C^{||(1)} ({\bf x})  \  , \\
&& \bar  D_{ij} \chi^{\parallel(1)} =
    D_{ij} \chi^{\parallel(1)}
    -2  D_{ij} C^{||(1)}({\bf x})  \  , \label{phigauge2}
\ea
\be  \label{gaugetrchiT}
\bar \chi^{\top (1)}_{ij} = \chi^{\top (1)}_{ij} \ .
\ee
We   shall first give the 2nd-order gauge transformations
for  the scalar-tensor coupling.
From the general formulas (C43), (C48),  and (C49) of Ref.~\cite{ScalarScalar},
keeping only the   $\chi^{\top(1)}_{ij} $-linear-dependent terms
and using  the conditions  (\ref {U10}) and  (\ref {U20}),
the 2nd-order vector field  $\xi^{(2)\mu } $ is given
as the following
\be\label{alpha2st2}
\alpha^{(2)}
=\frac{ A^{(2)}}{\tau^2} \, ,
\ee
\be\label{beta2st2}
\beta^{(2)}_{,\,i}
= C^{||(2)} (\mathbf x)_{,\,i}   \,,
\ee
\be\label{d2st2}
d^{(2)}_i
=C^{\perp(2)}_{\,i}(\mathbf x) \  ,
\ee
where   $A^{(2)}$ is an arbitrary constant,
$C^{||(2)}(\mathbf x)$ is an arbitrary function,
$C^{\perp(2)}_{\,i}(\mathbf x)$ is an arbitrary curl vector;
all of them shall be linearly  depending on   $\chi^{\top(1)}_{ij}$
at some fixed time.
Accordingly, by the general  formulas
(C50), (C56)--(C58) in Ref.~\cite{ScalarScalar}
of the 2nd-order residual gauge transformation of metric perturbations,
 keeping only the scalar-tensor terms, one obtains:
\be\label{piiSt2Trans2}
\bar \phi^{(2)}_{s(t)}=
\phi^{(2)}_{s(t)}
+\frac23 C^{||(1),kl}\chi^{\top(1)}_{kl}
+\frac{2 }{\tau^3}A^{(2)}
+ \frac13\nabla^2C^{||(2)}
,
\ee
\bl\label{chi||2trans4}
D_{ij}\bar\chi^{||(2)}_{s(t)}
= & D_{ij}\chi^{||(2)}_{s(t)}
-D_{ij}\nabla^{-2}\nabla^{-2}\big[
9C^{||(1),k lm}\chi^{\top(1)}_{kl,m}
+  6\chi^{\top(1)}_{kl}\nabla^2 C^{||(1),kl}
\big] \nn \\
&
+2 D_{ij}\nabla^{-2}\big[
 C^{||(1),kl}\chi^{\top(1)}_{kl} \big]
- 2D_{ij}C^{||(2)} ,
\el
\bl\label{chiPerp2Trans33}
\bar\chi^{\perp(2)}_{s(t)ij}
=&\chi^{\perp(2)}_{s(t)ij}
-2 \partial_i\nabla^{-2}\big(
  \chi^{\top(1)}_{kj}\nabla^2C^{||(1),k}
+2\chi^{\top(1)}_{kj,l}C^{||(1),kl}
+C^{||(1),kl}_{,j}\chi^{\top(1)}_{kl}
\big)
\nn\\
&
+\partial_i\partial_j\nabla^{-2}\nabla^{-2}\big(
 4\chi^{\top(1)}_{kl}\nabla^2C^{||(1),kl}
+6 C^{||(1),klm}\chi^{\top(1)}_{kl,m}
\big)
-  C^{\perp(2)}_{i,j}
\nn \\
&
+ ( i  \leftrightarrow  j ),
\el
\bl\label{chiT2Trans4}
\bar\chi^{\top(2)}_{s(t)ij}
=&\chi^{\top(2)}_{s(t)ij}
-\bigg[\frac{8}{\tau^3}A^{(1)}\chi^{\top(1)}_{ij}
+\frac{2}{\tau^2}A^{(1)}\chi^{\top(1)\,'}_{ij}
+2 C^{||(1),k}\chi^{\top(1)}_{ij,k}\bigg]
\nn\\
&
+\nabla^{-2}\bigg[
-2C^{||(1),k}_{,\,i}\nabla^2\chi^{\top(1)}_{kj}
-2C^{||(1),k}_{,\,j}\nabla^2\chi^{\top(1)}_{ki}
+2\chi^{\top(1)}_{ki,j}\nabla^2C^{||(1),k}
\nn\\
&
+2\chi^{\top(1)}_{kj,i}\nabla^2C^{||(1),k}
+4C^{||(1),kl}\chi^{\top(1)}_{ki,jl}
+4C^{||(1),kl}\chi^{\top(1)}_{kj,il}
+2C^{||(1),kl}_{,ij}\chi^{\top(1)}_{kl}
\nn\\
&
-2C^{||(1),kl}\chi^{\top(1)}_{kl,ij}
+2C^{||(1),kl}\nabla^2\chi^{\top(1)}_{kl}\delta_{ij}
+C^{||(1),klm}\chi^{\top(1)}_{kl,m}\delta_{ij}
\bigg]
\nn\\
&
-\partial_i\partial_j\nabla^{-2}\nabla^{-2}\bigg[
2\chi^{\top(1)}_{kl}\nabla^2C^{||(1),kl}
+3 C^{||(1),klm}\chi^{\top(1)}_{kl,m}\bigg] \ .
\el
where   the  constants
$A^{(1)}$,
$C^{||(1)}(\mathbf x)$,
$C^{\perp(1)}_{\,i}(\mathbf x)$
are all independent of  the tensor  $\chi^{\top(1)}_{ij}$.
As   Eq.(\ref {chiT2Trans4}) tells,
the transformation of 2nd-order tensor
involves  only  the vector field  $\xi^{(1)}$,
independent of $\xi^{(2)}$.

It should be pointed out that
the roles of $\xi^{(1)}$ and $\xi^{(2)}$ are different.
When  one sets $\xi^{(2)}=0$
in Eq.(\ref{xmutransf})  \cite{ScalarScalar,Abramo1997},
only  $\xi^{(1)}  $ remains,
which ensures $\bar g^{(1)}_{00}=0$,  $\bar g^{(1)}_{0i}=0$.
Nevertheless, now one has no freedom
to make  $\bar g^{(2)}_{00}=0$ and  $\bar g^{(2)}_{0i}=0$,
 since  $\xi^{(1)}$ has  already  been used
in obtaining $\bar g^{(1)}_{00}=0$ and  $\bar g^{(1)}_{0i}=0$
and keeping the obtained 1st-order perturbations unchanged
in the fixed 1st-order synchronous coordinate.
Thus, 2nd-order transformations from synchronous to synchronous
can not effectively be made when one sets  $\xi^{(2)}=0$.
On the other hand,
if one does not  transform the 1st order,
but only   transforms
the 2nd-order metric perturbations \cite{Gleiser1996},
one simply sets $\xi^{(1)}=0$ but $\xi^{(2)} \ne 0$.
Then
(\ref {piiSt2Trans2})--(\ref{chiT2Trans4}) reduce to
\be\label{piiSt2Trans3}
\bar \phi^{(2)}_{s(t)}=
\phi^{(2)}_{s(t)}
+\frac{2 }{\tau^3}A^{(2)}
+   \frac13\nabla^2C^{||(2)},
\ee
\be\label{chi||2trans5}
D_{ij}\bar\chi^{||(2)}_{s(t)}
=
D_{ij}\chi^{||(2)}_{s(t)}
-  2D_{ij}C^{||(2) },
\ee
\be\label{chiPerp2Trans4}
\bar\chi^{\perp(2)}_{s(t)ij}
=\chi^{\perp(2)}_{s(t)ij}
- \bigg(C^{\perp(2)}_{i,j}
+C^{\perp(2)}_{j,i}
\bigg),
\ee
\be\label{chiT2Trans44}
\bar\chi^{\top(2)}_{s(t)ij}
=\chi^{\top(2)}_{s(t)ij} .
\ee
From the transformation formula
$\bar\rho ^{(2)}_{s(t)} = \rho ^{(2)}_{s(t)} - \mathcal{L}_{\xi^{(2)}} \rho^{(0)}$ \cite{ScalarScalar},
the transformation of the 2nd-order density perturbation is
\be
\bar\rho   ^{(2)}_{s(t)}  =   \rho  ^{(2)}_{s(t)} +6 \frac{A^{(2)}} {\tau^3}  \rho^{(0)}  \, ,
\ee
by which the transformation of the 2nd-order density contrast  is given by
\be \label{trfmdelta2}
\bar \delta   ^{(2)}_{s(t)}  =   \delta   ^{(2)}_{s(t)} +6   \frac{A^{(2)}} {\tau^3} \, .
\ee
where $A^{(2)}$ shall be linear  depending on   $\chi^{\top(1)}_{ij}$
at some fixed time.
These transformations by $\xi^{(2)}$
have the same structure as the 1st-order gauge transformations \cite{ScalarScalar}.

By this  result, we can identify the residual gauge modes in the 2nd-order solutions
for the scalar-tensor coupling.
In  the solution of scalar $\phi^{(2)}_{s(t)}$ in (\ref{phi2st}),
the constant terms
$(\phi^{(2)}_{s(t)\,0} -\frac{1}{3}\delta_{s(t)\,0}^{(2)}
-\frac{\tau_0^2}{9}\varphi^{,\,ij}\chi^{\top(1)}_{0ij} )$
are a residual gauge mode,
which can be changed  by a  choice of $C^{||(2)}$.
In the solution of scalar $D_{ij}\chi^{\parallel(2)\, }_{s(t)}$ in (\ref{chi112qqq}),
 the constant term $D_{ij}Z  $ is also a gauge term
 that will be changed by  $C^{||(2)}$ accordingly.
Similarly,
in the solution of vector  $\chi^{\perp(2)}_{s(t)ij}$ in (\ref{chiSTperp2}),
the  constant term
 $\nabla^{-2} (Q_{i,j} +Q_{j,i}) $
is a residual gauge mode
and can be removed by a choice of $(C^{\perp(2)}_{i,j} + C^{\perp(2)}_{j,i})$,
but other time-dependent terms  in (\ref{chiSTperp2})
are not gauge modes.
In contrast,
the 2nd-order tensor is  invariant under the transformation by $\xi^{(2)}$
as  demonstrated by Eq.(\ref {chiT2Trans44}),
and
the solution of tensor in  Eq.(\ref{solgw})  thus contains no gauge mode.

Next consider the case of  tensor-tensor coupling,
the analysis is  similar to the above paragraphs.
In particular,
the 2nd-order residual gauge transformation
is effectively implemented only by the 2nd-order  vector field $\xi^{(2)}$ even given nonzero $\xi^{(1)}$,
and the gauge transformations are similar to
(\ref {piiSt2Trans3})--(\ref {chiT2Trans44}) and (\ref {trfmdelta2})
\be\label{piiT2Trans3}
\bar \phi^{(2)}_{T}=
\phi^{(2)}_{T}
+\frac{2 }{\tau^3}A^{(2)}
   + \frac13\nabla^2C^{||(2)}(\mathbf x)
\ee
\be\label{chi||2Ttrans5}
D_{ij}\bar\chi^{||(2)}_{T}
=
D_{ij}\chi^{||(2)}_{T}
   - 2D_{ij}C^{||(2) }(\mathbf x)
\ee
\be\label{chiPerp2TTrans4}
\bar\chi^{\perp(2)}_{T  ij}
=\chi^{\perp(2)}_{ T ij}
- \bigg(C^{\perp(2)}_{i,j}(\mathbf x)
+C^{\perp(2)}_{j,i}(\mathbf x)
\bigg)
\ee
\be\label{chiT2TTrans44}
\bar\chi^{\top(2)}_{T  ij}
=\chi^{\top(2)}_{T ij}
\ee
\be \label{trfmdelta2T}
\bar \delta   ^{(2)}_{T}
  =   \delta   ^{(2)}_{T} +6   \frac{A^{(2)}} {\tau^3} \, .
\ee
where
$A^{(2)}$,
$C^{||(2)}(\mathbf x)$,
$C^{\perp(2)}_{\,i}(\mathbf x)$
shall  depend on  tensor-tensor terms such as  $\chi^{\top(1)}_{ij} \chi^{\top(1)}_{kl }$
at some fixed time.
By   (\ref{piiT2Trans3})--(\ref{chiT2TTrans44}),
the residual gauge modes in the  solutions of 2nd-order metric perturbations
for the tensor-tensor coupling can be identified similarly.
For instance,
the constant terms in  the solutions (\ref {phi2T}), (\ref {chi112qqqTT}), and (\ref{chiTTperp2})
are residual gauge modes,
and can be changed by choices of
 $C^{||(2)} $ and  $C^{\perp(2)}_{\,i} $ respectively.
Furthermore,
Eq.(\ref {chiT2TTrans44}) shows that
 $\chi^{\top(2)}_{ T  i j}$
generated by the tensor-tensor coupling
is  gauge invariant,
so that the solution of (\ref{solgwTT}) in synchronous coordinates
contains no gauge mode.

  \section{Conclusion}

We have studied  the  2nd-order cosmological
 perturbations in the Einstein-de Sitter Universe
in synchronous coordinates.
The scalar-tensor and tensor-tensor types of couplings of 1st-order metric perturbations
serve as a part of effective source for the 2nd-order metric perturbations.
For each coupling, respectively,
the 2nd-order  perturbed Einstein equation has been solved
with general initial conditions,
and   the explicit solutions
of scalar, vector, and tensor 2nd-order metric perturbations
 have been obtained.

We have also performed  general 2nd-order  synchronous-to-synchronous  gauge transformations,
which are generated by a 1st-order vector field
and a 2nd-order vector field.
For the scalar-tensor and tensor-tensor couplings respectively,
we have identified all the residual gauge modes of
the 2nd-order metric perturbations in synchronous coordinates.
By analysis, we point out that, holding  the 1st-order solutions fixed,
only the 2nd-order transformation vector field
is effective in carrying out the 2nd-order transformations.
This is because of the fact that
the 1st-order vector field  has been already determined
in the 1st-order transformations.
In particular,
the 2nd-order  tensor is found to be invariant
under 2nd-order gauge transformations
just like the 1st-order tensor is invariant
under the 1st-order  transformations.

Thus, together with the case of scalar-scalar couplings in our previous work,
we have obtained  the  full solution of the 2nd-order cosmological perturbations
and all their  residual gauge modes
of the Einstein-de Sitter Universe in synchronous coordinates,
where all  the couplings of 1st-order perturbations are included.
As a possible application of the results of 2nd-order perturbations to CMB,
one can use the derived expressions
$\gamma_{ij}^{(1)} + \frac{1}{2} \gamma_{ij}^{(2)}$
into the Sachs-Wolfe term of the Boltzmann equation of photon gas.
The corresponding spectra $C_l^{XX}$ of on CMB anisotropies and polarization
will contain the contributions from  the 2nd-order   effects of $\gamma_{ij}^{(2)}$.

\

\textbf{Acknowledgements}

Y. Zhang is supported by
NSFC Grant No. 11275187,
No. NSFC 11421303, No. 11675165, No. 11633001,
SRFDP, and CAS, the Strategic Priority Research Program
``The Emergence of Cosmological Structures"
of the Chinese Academy of Sciences, Grant No. XDB09000000.

\end{document}